\documentclass[fleqn,usenatbib]{mnras}

\usepackage{amsmath}

\usepackage[T1]{fontenc}

\usepackage{caption}
\usepackage{subcaption}
\usepackage{rotating}
\usepackage{longtable}
\usepackage{lscape}
\DeclareRobustCommand{\VAN}[3]{#2}
\let\VANthebibliography\thebibliography
\def\thebibliography{\DeclareRobustCommand{\VAN}[3]{##3}\VANthebibliography}

\usepackage{graphicx}	
\usepackage{amsmath}	
\usepackage{amssymb}	
\makeatletter
\DeclareRobustCommand{\ion}[2]{%
  \text{#1\,\check@mathfonts\fontsize\sf@size\z@\selectfont #2}%
}
\makeatother

\newcommand{\HII}{\ion{H}{II}}
\newcommand{\NII}{\ion{N}{II}}
\newcommand{\OII}{\ion{O}{II}}
\newcommand{\OIII}{\ion{O}{III}}
\newcommand{\SII}{\ion{S}{II}}
\newcommand{\SIII}{\ion{S}{III}}

\newcommand{\ha}{\textup{H\ensuremath{\alpha}}}
\newcommand{\hb}{\textup{H\ensuremath{\beta}}}
\usepackage{newtxtext,newtxmath}

\title[Stellar feedback in NGC~300]{The impact of pre-supernova feedback and its dependence on environment}

\author[A. F. McLeod et al.]{Anna F. McLeod,$^{1,2}$\thanks{E-mail: anna.mcleod@durham.ac.uk} Ahmad A. Ali,$^{3}$ M{\'e}lanie Chevance,$^{4}$ Lorenza Della Bruna,$^{5}$ \newauthor Andreas Schruba,$^{6}$ Heloise F. Stevance,$^{7}$ Angela Adamo,$^{5}$ J. M. Diederik Kruijssen,$^{4}$  \newauthor  Steven N.~Longmore,$^{8}$ Daniel R. Weisz,$^{9}$ and Peter Zeidler$^{10}$
\\
$^{1}$Centre for Extragalactic Astronomy, Department of Physics, Durham University, South Road,  Durham DH1 3LE, UK\\
$^{2}$Institute for Computational Cosmology, Department of Physics, University of Durham, South Road, Durham DH1 3LE, UK\\
$^{3}$Department of Physics and Astronomy, University of Exeter, Stocker Road, Exeter EX4 4QL, United Kingdom\\
$^{4}$Astronomisches Rechen-Institut, Zentrum f{\"u}r Astronomie der Universit{\"a}t Heidelberg, M{\"o}nchhofstra\ss e 12-14, D-69120 Heidelberg, Germany\\
$^{5}$Department of Astronomy, Oskar Klein Centre, Stockhom University, AlbaNova University Centre, 106 91 Stockholm, Sweden\\
$^{6}$Max-Planck-Institut f{\"u}r extraterrestrische Physik, Giessenbachstra\ss e 1, D-85748 Garching\\
$^{7}$University of Auckland, Department of Physics and Astronomy, 38 Princes Street, 1142, Auckland, New Zealand\\
$^{8}$Astrophysics Research Institute, Liverpool John Moores University, Liverpool, L3 5RF, UK\\
$^{9}$Department of Astronomy, University of California Berkeley, Berkeley, CA 94720, USA\\
$^{10}$AURA for the European Space Agency (ESA), Space Telescope Science Institute, 3700 San Martin Drive, Baltimore, MD 21218, USA
}

\date{Accepted 2021 September 17. Received 2021 September 17; in original form 2021 July 20.
}

\pubyear{2015}

\begin{document}
\label{firstpage}
\pagerange{\pageref{firstpage}--\pageref{lastpage}}
\maketitle

\begin{abstract}
Integral field units enable resolved studies of a large number of star-forming regions across entire nearby galaxies, providing insight on the conversion of gas into stars and the feedback from the emerging stellar populations over unprecedented dynamic ranges in terms of spatial scale, star-forming region properties, and environments.  We use the VLT/MUSE legacy data set covering the central $35$~arcmin$^{2}$ (${\sim}12$~kpc$^{2}$) of the nearby galaxy NGC~300 to quantify the effect of stellar feedback as a function of the local galactic environment. We extract spectra from emission line regions identified within dendrograms, combine emission line ratios and line widths to distinguish between \HII\ regions, planetary nebulae, and supernova remnants, and compute their ionised gas properties, gas-phase oxygen abundances, and feedback-related pressure terms. For the \HII\ regions, we find that the direct radiation pressure ($P_\mathrm{dir}$) and the pressure of the ionised gas ($P_{\HII}$) weakly increase towards larger galactocentric radii, i.e.\ along the galaxy's (negative) abundance and (positive) extinction gradients. While the increase of $P_{\HII}$ with galactocentric radius is likely due to higher photon fluxes from lower-metallicity stellar populations, we find that the increase of $P_\mathrm{dir}$ is likely driven by the combination of higher photon fluxes and enhanced dust content at larger galactocentric radii. In light of the above, we investigate the effect of increased pre-supernova feedback at larger galactocentric distances (lower metallicities and increased dust mass surface density) on the ISM, finding that supernovae at lower metallicities expand into lower-density environments, thereby enhancing the impact of supernova feedback.  
\end{abstract}




\defcitealias{mcleod20}{M20}

\section{Introduction}
\label{sec:intro}

Stellar feedback is a multi-scale phenomenon, arising from small (pc) scales of the feedback-driving stars and their natal clouds but having profound effects up to the (kpc) scales of entire galaxies. Via a series of different mechanisms (i.e.\ protostellar outflows, radiation pressure, ionisation, stellar winds, and supernovae; see e.g.\ \citealt{krumholz14a,girichidis20}), the feedback generated during the lives and deaths of stars above ${\sim}8$~M$_{\odot}$ enriches the interstellar medium \citep[ISM; e.g.][]{scannapieco06,maiolino19,agertz20}, drives the expansion of \HII\ regions and the disruption of molecular clouds \citep[e.g.][]{krumholz06,chevance20c}, regulates the formation of star clusters \citep[e.g.][]{ostriker10,kruijssen12,krumholz19}, controls the baryon cycle in star-forming galaxies \citep[e.g.][]{mckee77,hopkins11,naab17}, reshapes the dark matter distributions of dwarf galaxies \citep[e.g.][]{governato2010,trujillogomez21}, and facilitates the dispersal of proto-planetary discs \citep[e.g.][]{johnstone98,scally01}.

Many numerical studies have shown that including stellar feedback in galaxy simulations is required to recover global observational properties such as star formation rates and star formation efficiencies \citep[e.g.][]{agertz13,hopkins14,crain15,fujimoto19,keller21}, to replicate well known scaling relations like the Kennicutt--Schmidt relation \citep{schmidt59,kennicutt98}, and to reconcile the long global depletion time (${\sim}1$~Gyr; e.g.\ \citealt{bigiel08,leroy08}) with the short gravitational collapse time-scale (${\sim}10$~Myr; e.g.\ \citealt{heyer15,utomo18,schruba19}). Different numerical studies implement different (combinations of) feedback mechanisms, and where multiple ones are included, efforts go towards understanding the relative importance of individual feedback mechanisms. For example, supernovae~(SNe) have long thought to be the dominating source of feedback, but both simulations \citep[e.g.][]{haid18,lucas20,semenov21,keller21b} and observations \citep[e.g.][]{kruijssen19,chevance20b} now clearly show that early stellar feedback (i.e.\ radiation pressure, ionisation, and stellar winds) play a major role in regulating the impact of SN feedback by altering the conditions of the interstellar medium (ISM) prior to the first SN events. These results indicate that not only is the environment being affected by stellar feedback, but environmental properties in turn set the effectiveness of feedback. While the impact of feedback on the environment is an established area of active research, the impact of the environment on feedback is now starting to be explored from the observational perspective \citep[e.g.][]{lopez11,lopez14,chevance16,barnes20,olivier20}. Outstanding questions include: What are the dominant feedback mechanisms from massive stars as a function of their stellar properties (e.g.\ mass, chemical composition, rotation rate, binarity, evolutionary phase)? How does feedback change with environment (e.g.\ metallicity, ambient gas density, location within a galaxy)? How does our knowledge of feedback change with physical scale, from small (clouds) to large (galaxies) scales? 

Over the past decade, the increasing availability of large field-of-view, large wavelength coverage, and medium spectral resolution integral field unit (IFU) instruments has enabled the simultaneous study of the feedback-driving stellar populations and the feedback-affected matter. For example, this has led to optical studies of the stellar and ionised gas properties and kinematics of entire spatially-resolved star-forming regions in the Milky Way \citep[e.g.][]{m16,weilbacher15,pillars,flagey20}, the Magellanic Clouds \citep[e.g.][]{castro18,mcleod18b}, and nearby galaxies \citep[e.g.][]{monreal11,monreal12,westmoquette13,mcleod20}. While these regions are observationally convenient for detailed multi-wavelength studies of feedback on small scales or in select galactic hosts, they are not broadly representative of star formation and feedback over all parameter space and do not consider the effects of feedback in the larger context of their galactic hosts. The need for large region samples spanning a vast parameter space and being spatially-matched with available multi-wavelength ISM observations has produced large nearby galaxy IFU surveys such as SIGNALS\footnote{Star formation, ionised Gas and Nebular Abundances Legacy Survey with SITELLE.} \citep{laurie19} and PHANGS\footnote{Physics at High Angular Resolution in Nearby Galaxies.}-MUSE \citep[for early results see \citealt{kreckel19,pessa21}]{schinnerer19,emsellem21}. For tens of thousands of regions, these surveys deliver spatially-resolved (for the nearest systems up to a few~Mpc) and integrated (beyond a few~Mpc) stellar and ionised gas properties, enabling environmental studies of stellar feedback in orders of magnitude more star-forming regions than previously possible. These are also well-matched with recent advancements made in computational galaxy evolution models.

Here, we study the environmental dependence of ionised gas properties and feedback-related pressure terms in the nearby galaxy NGC~300, and study their implications in the context of early pre-SN and subsequent SN feedback. This galaxy has been the focus of two recent feedback-related studies upon which this present paper builds. (1) \citet{kruijssen19} use a novel statistical method based on the spatial decorrelation between young stars and molecular gas to infer feedback-related quantities across the galactic disc of NGC~300 (i.e.\ molecular cloud lifetimes, feedback time-scales, outflow velocities, mass-loading factors, and star formation efficiencies). They show that star formation in NGC~300 is rapid and inefficient, with giant molecular clouds (GMCs) having integrated star formation efficiencies of only $2{-}3$~per~cent, but being dispersed by feedback from massive stars within $1.5\pm0.2$~Myr. (2) \citet[][henceforth referred to as \citetalias{mcleod20}]{mcleod20} study 5~\HII\ regions in NGC~300 in a spatially resolved manner and find that their expansion is governed by the pressure of the ionised gas and by the winds from the massive stars within them. 

In this paper, we use a legacy value data set covering the inner star-forming disc of NGC~300 taken with the VLT/MUSE instrument \citep{muse}, consisting of a contiguous $7\arcmin \times 5\arcmin$ mosaic (${\sim}4~\mathrm{kpc} \times 3~\mathrm{kpc}$) and covering the galaxy out to galactocentric radii of about $0.45 R_{25}$ (with $R_{25} \sim 5.33$~kpc being the optical radius; \citealt{paturel03}). NGC~300 is an ideal target to study stellar feedback: it is the closest \citep[$D \sim 2$~Mpc;][]{dalcanton09}, non-interacting, star-forming disc galaxy that can be mapped at the necessary spatial resolution (i.e.\ $1\arcsec$, which corresponds to ${\sim}10$~pc, resolving individual star-forming regions and supernova remnants). The large spatial coverage (to cover most of the star-forming disc) available not only in the optical with MUSE but throughout the electromagnetic spectrum \citep[e.g.][]{helou04,westmeier11,riener18,kruijssen19,schruba19} makes NGC~300 the ideal target for simultaneous resolved feedback, stellar population, and ISM studies. Closer galaxies like the Magellanic Clouds, M31 or M33 do not allow similar large-scale multi-wavelength mapping due to their large angular sizes, while more distant galaxies (beyond a few Mpc) do not allow multi-wavelength studies with similar spatial resolution across the optical, infrared, \mbox{mm/sub-mm}, and radio. NGC~300 perfectly bridges between ${\sim}100$~pc resolution IFU surveys of nearby galaxies like PHANGS, and upcoming \mbox{(sub-)\,pc} scale resolution IFU surveys of the Milky Way and the Magellanic Clouds like \mbox{SDSS-V/LVM} \citep{kollmeier17}.
Further, NGC~300 offers a favorable inclination of ${\sim}40^{\circ}$ \citep{puche90}, it is actively forming stars at a rate between ${\sim}0.08$ and ${\sim}0.30$~M$_{\odot}$~yr$^{-1}$ \citep[][and references therein]{kang16}, and has a well-studied population of \HII\ regions \citep[e.g.][]{deharveng88,bresolin09,faesi14}, planetary nebulae \citep[PNe; e.g.][]{soffner96,pena12,stasinska13}, and supernova remnants \citep[SNRs; e.g.][]{blair97,millar12,vucetic15}.

This paper is organised as follows. After a brief overview of the VLT/MUSE observations in Section~\ref{sec:data}, in Section~\ref{sec:regions}, we describe the methods used to identify and classify emission line regions. In Section~\ref{sec:gas}, we compute ionised gas properties and feedback-related pressure terms for the detected \HII\ regions and discuss environmental dependencies. In Section~\ref{sec:preshock}, we analyse the environment in which the covered SN events occurred. Finally, we summarise our findings and conclusions in Section~\ref{sec:conclusion}.

\section{Observations}
\label{sec:data}

\begin{figure*}
    \centering
    \includegraphics[scale=0.85]{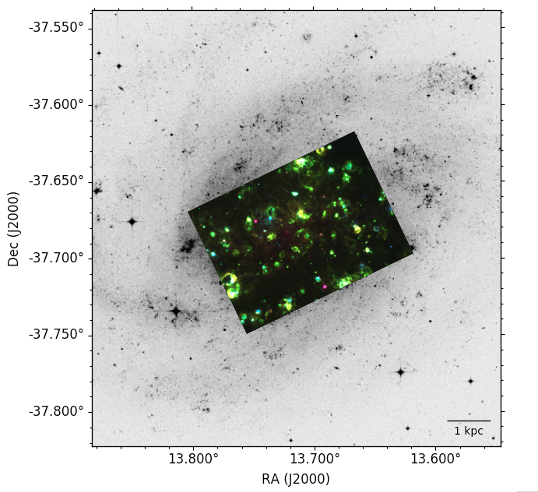}
    \caption{ESO-DSS image of NGC~300 (gray scale) with the continuum-subtracted RGB composite of the 35-pointing MUSE mosaic overlayed (red = $[\SII]\lambda6717$, green = \ha, blue = $[\OIII]\lambda5007$). The size of the mosaic is $7\arcmin \times 5\arcmin$, thus covering the inner (${\sim}4~\mathrm{kpc} \times 3~\mathrm{kpc}$) of the galaxy.}
    \label{fig:rgb}
\end{figure*}

This work is based on the VLT/MUSE data set of NGC~300 first presented in \citetalias{mcleod20}. The data were taken in the nominal wavelength range of the MUSE instrument (${\sim} 4750{-}9350$~\AA) and using its wide-field mode (${\sim}1\arcmin \times 1\arcmin$ per pointing), as part of the observing program \mbox{098.B-0193(A)} (PI McLeod). 
As opposed to \citetalias{mcleod20}, where only~2 of the NGC~300 MUSE data cubes are analysed, here we exploit the full coverage of the in total 35 individual mosaic pointings which cover a $7\arcmin \times 5\arcmin$ contiguous mosaic of the central star-forming disc of NGC~300.
The data were taken prior to the availability of the Adaptive Optics system for MUSE's large field-of-view, and seeing-limited angular resolutions in a range of about $0\farcs45 {-} 1\farcs3$ were achieved (see Table~\ref{tab:cubes}).  Each individual mosaic pointing was observed three times in a $90\degr$-rotation dither pattern with an exposure time of $900$~seconds per rotation. The full mosaic is shown in Fig.~\ref{fig:rgb}, which consists of a three-colour composite of the 35~pointings with three emission lines tracing the ionised gas (red = $[\SII]\lambda6717$, green = \ha, blue = $[\OIII]\lambda5007$), overlayed on an optical ESO-DSS image for reference.
Observational details of the individual pointings are given in Table~\ref{tab:cubes}. 

As described in \citetalias{mcleod20}, we proceed in reducing the data with the MUSE pipeline \citep{pipeline} in the {\sc esorex} environment with the standard static calibration files. For each specific observing block we use the available calibration files from the ESO archive, and subtract the sky lines according to \citet{zeidler19} (for brevity we do not describe sky subtraction details here and refer the interested reader to \citetalias{mcleod20}). The flux calibration (also performed with the MUSE pipeline) is done for each observing block using the matched nightly standard star observations.
The three exposures of each individual pointing are combined into single cubes with the built-in exposure combination recipes of the MUSE pipeline. 

\begin{table}
\begin{center}
\caption{Observational information of the 35 mosaic pointings obtained with the  VLT/MUSE instrument for NGC~300. See text in Section~\ref{sec:data}.}
\begin{tabular}{lccc}
\hline 
\hline
Field & Field centre & Observation date & Seeing\\
\hline
 & (J2000) & (YYYY-MM-DD) & (arcsec) \\
\hline
H1 & 00:54:59.83 -37:39:42.0 & 2016-10-01 & 0".82 \\
H2 & 00:54:55.40 -37:39:17.0 & 2016-10-01 & 0".96\\
H3 & 00:54:50.99 -37:38:51.8 & 2016-10-01 & 0".68\\
H4 & 00:54:46.55 -37:38:26.7 & 2016-10-04 & 1".29\\
H5 & 00:55:06.51 -37:41:25.5 & 2016-10-05 & 1".28\\
H6 & 00:55:02.08 -37:41:00.9 & 2016-10-05 & 1".07\\
H7 & 00:54:57.65 -37:40:35.7 & 2016-10-05 & 1".06\\
H8 & 00:54:53.22 -37:40:10.3 & 2016-10-05 & 0".84\\
H9 & 00:54:48.81 -37:39:45.4 & 2016-11-07 & 0".46\\
H10 & 00:54:44.37 -37:39:20.3 & 2016-11-08 & 0".82\\
H11 & 00:54:39.95 -37:38:55.1 & 2016-11-08 & 0".69\\
H12 & 00:55:04.35 -37:42:19.4 & 2016-11-08 & 0".70\\
H13 & 00:54:59.90 -37:41:54.7 & 2016-11-08 & 0".60\\
H14 & 00:54:55.47 -37:41:29.3 & 2016-11-08 & 0".98\\
H15 & 00:54:57.70 -37:42:48.2 & 2016-11-08 & 1".00\\
H16 & 00:54:55.52 -37:43:42.0 & 2016-12-19 & 0".63\\
H17 & 00:54:51.08 -37:43:16.8 & 2016-12-23 & 1".07\\
L1 & 00:55:08.69 -37:40:32.3 & 2016-12-23 & 1".11\\
L2 & 00:55:04.26 -37:40:06.8 & 2016-12-23 & 0".67\\
L3 & 00:54:42.13 -37:38:01.3 & 2016-12-24 & 1".08\\
L4 & 00:54:51.04 -37:41:04.2 & 2016-12-26 & 0".92\\
L5 & 00:54:46.63 -37:40:38.9 & 2017-01-02 & 1".29\\
L6 & 00:54:42.19 -37:40:13.7 & 2017-01-02 & 1".01\\
L7 & 00:54:37.77 -37:39:48.5 & 2017-01-04 & 1".32\\
L8 & 00:55:02.15 -37:43:13.1 & 2018-07-03 & 1".06\\
L9 & 00:54:53.27 -37:42:22.6 & 2017-01-05 & 0".87\\
L10 & 00:54:48.84 -37:41:57.9 & 2017-01-06 & 0".53\\
L11 & 00:54:44.43 -37:41:32.7 & 2017-01-06 & 0".44\\
L12 & 00:54:39.99 -37:41:07.4 & 2017-01-06 & 0".52\\
L13 & 00:54:35.57 -37:40:42.2 & 2017-01-07 & 0".42\\
L14 & 00:54:59.95 -37:44:06.8 & 2017-01-07 & 0".57\\
L15 & 00:54:46.64 -37:42:51.6 & 2017-01-16 & 0".51\\
L16 & 00:54:42.21 -37:42:26.4 & 2017-01-27 & 0".78\\
L17 & 00:54:37.79 -37:42:00.9 & 2018-07-04 & 0".85\\
L18 & 00:54:33.37 -37:41:36.2 & 2018-07-04 & 0".77\\
\hline
\hline
\label{tab:cubes}
\end{tabular}
\end{center}
\end{table}

The extended nature of the types of objects analysed here (i.e.\ mainly \HII\ regions and SNRs) necessarily means that some of these significantly overlap between different cubes, making it imperative to mosaic these cubes such that an integrated spectrum for the regions that overlap between two or more individual pointings can be extracted. Two issues arise prior to producing a mosaic of the cubes. First, combining 35~individual MUSE cubes, each several GBs in size, would result in a single, giant cube of unreasonably large (and thus unwieldy) file size. Second, the individual pointings were observed during different nights and at varying observing conditions, leading to relative flux offsets across the field-of-view. To overcome these two problems we proceed in the following way:

\begin{itemize}
    \item The 35~cubes are first resampled to a common wavelength grid (to overcome slight wavelength offsets between cubes) and divided into sub-cubes spanning 500~wavelength elements each (i.e.\ $625$~\AA; the full wavelength range of each MUSE cube is $4750{-}9350$~\AA\ and the sampling is $1.25$~\AA).
    \item Individual 2D slices from each sub-cube are then mosaicked using the {\sc Astropy} \citep{astropy:2018} Montage wrapper with a background match.
    \item The 2D mosaics are then recombined into data cubes which now cover the entire field-of-view and span across a manageable wavelength range.
\end{itemize}

Spectra of regions of interest can then be extracted from each one of the large cubes spanning about $625$~\AA\ each, and combined to cover the entire wavelength range. To assess the performance of combining the cubes as described, we compare fluxes obtained from the mosaicked MUSE data to those reported in previous spectroscopic studies of regions in NGC~300While the comparison with fluxes from photometric studies \citep[e.g.,][]{faesi14} is feasible, it requires carefully reproducing the filter parameters of the used instrument. The \HII\ regions used for this comparison are spread across the MUSE FOV to compare fluxes across different pointings. We compare MUSE fluxes with those reported in \citet[][]{toribio16}, who use VLT/UVES data to derive abundances of 7 \HII\ regions in NGC~300, 3 of which overlap with the multi-night MUSE data (specifically, their R20, R23, and R76a are in MUSE fields H8, H12, and L4, respectively). Crucially, for each of these regions, \citeauthor{toribio16} report the central coordinates and area covered by their observations (their Table 1), and we extract integrated spectra from the MUSE data accordingly. The comparison, summarised in Table \ref{tab:flux_comp}, shows excellent agreement (within errors) between the MUSE and UVES fluxes, with the exception of the \ha\ flux of R23, which is likely due to a repetition typo in Table 1 of \citeauthor{toribio16}. We note that for the purpose of this paper, we are only showing the comparison for \ha, \hb, and [\NII]$\lambda$6584. We do not attempt a comparison with fluxes from \citet[][]{bresolin09} or \citet[][]{stasinska13}, as both of these studies use VLT/FORS2 data but do not specify the slit lengths adopted for individual regions, thus hindering a direct comparison. Further, for the purpose of the subsequent analysis, we note that when making the mosaic no PSF matching was performed, as the region size constraints described below ensure that all analysed objects are resolved regardless of the seeing achieved in a particular field, and because all analyses are performed on integrated spectra (and thus do not retain PSF information).

\begin{table*}
    \centering
    \begin{tabular}{llcccccc}
    \hline 
    \hline
         id & ID & F(\ha)$_{\mathrm{MUSE}}$ & F(\ha)$_{\mathrm{UVES}}$ & F([NII])$_{\mathrm{MUSE}}$ & F([NII])$_{\mathrm{UVES}}$ & F(\hb)$_{\mathrm{MUSE}}$ & F(\hb)$_{\mathrm{UVES}}$\\
         (this work) & (TSC16) &  & & & &  & \\
         \hline
          620 & R20 & 16.6$\pm$0.2 & 16.6$\pm$1.0 & 1.7$\pm$0.1 & 1.7$\pm$0.1 & 6.5$\pm$0.2 & 5.7$\pm$0.4 \\
          280 & R23 & 13.8$\pm$0.1 & 16.3$\pm$0.9 & 1.4$\pm$0.0 & 1.6$\pm$0.1 & 5.3$\pm$0.1 & 5.6$\pm$0.3 \\
          433 & R76a & 5.3$\pm$0.1 & 5.2$\pm$0.8 & 1.4$\pm$0.1 & 1.4$\pm$0.2 & 2.1$\pm$0.1 & 1.8$\pm$0.1\\
          \hline
    \end{tabular}
    \caption{Flux comparison for \HII\ regions in NGC~300 observed with MUSE (this work) and UVES \citep[][]{toribio16}. The first column refers to the dendrogram id (see Table \ref{tab:hii}), while the second column corresponds to the region id from \citeauthor{toribio16} (for central coordinates and region areas please refer to Table 1 in \citeauthor{toribio16}). The last six columns correspond to the reddening corrected \ha\, [\NII]$\lambda$6584, and \hb\ fluxes obtained from this study and \citeauthor{toribio16}, respectively. Fluxes are expressed in units of 10$^{-14}$ erg cm$^{-2}$ s$^{-1}$.}
    \label{tab:flux_comp}
\end{table*}

Emission line maps (such as those shown in Fig.~\ref{fig:rgb}) are obtained by collapsing the cubes over $\pm 3$~\AA\ (about $\pm 140$~km~s$^{-1}$ at \ha) around the lines of interest. The analysis described in the next section is based on a continuum-subtracted \ha\ map. For this, we first produce a continuum map by summing (i.e., collapsing) over a wavelength range equal in width to that used for the \ha\ line but covering a portion of nearby continuum (i.e.\ free of emission/\linebreak[0]{}absorption features, centered on $6540$~\AA), and then subtract this from the \ha\ map. The analyses presented here are not sensitive to small (${\sim}1\arcsec {-} 2\arcsec$) WCS shifts relative to, e.g.\ {\it HST} coordinates often observed in MUSE observations, and we therefore do not perform a WCS correction here. 
However, this is necessary when wanting to combine these with other data, and we note that data releases for this legacy MUSE data set of NGC~300 (which is part of several ongoing student projects) are scheduled to commence late 2022, in the form of fully reduced and WCS-corrected cubes, catalogs, and spectra.

\section{Region identification and classification}
\label{sec:regions}

While the objects of interest in this study are \HII\ regions and SNRs, the emission line regions as traced by the ionised gas also include sources of different nature (e.g.\ PNe, emission line stars, microquasars, ultra-luminous \mbox{X-ray} sources). It is therefore necessary to isolate the regions of interest from the general population of emission line sources. 

As already mentioned in Section~\ref{sec:intro}, NGC~300 is a well-studied galaxy and substantial catalogs of emission line regions have been compiled by previous studies \citep[e.g.][]{deharveng88, stasinska13, vucetic15}. This is the first time, however, that a large-scale IFU data set exists for this galaxy, meaning that we can now perform spectro-photo\-metric studies of these regions rather than relying on either only photometry, or targeted (slit) spectroscopy of selected regions. Given the rapid rise and wealth of already existing IFU observations of nearby galaxies, we use this MUSE data set of NGC~300 to explore new and efficient empirical identification and classification methods which can be readily adapted and applied to other optical IFU data of similar spatial resolution (e.g.\ the closest galaxies of the SIGNALS survey, \citealt{laurie19}, or NGC~7793, \citealt{dellabruna20,dellabruna21}).  

The MUSE NGC~300 data set gives access to simultaneous, spatially-resolved, photometric and spectroscopic information of ${>}100$ regions that are bright in the main nebular emission lines. We can therefore analyse population trends, such as radial abundance gradients, with improved number statistics. For example, the largest spectroscopic study of \HII\ region abundances in NGC~300 is that of \citet{bresolin09}, who obtained deep spectra for $28$~\HII\ regions. While our data are not as deep (e.g.\ we do not obtain sufficient signal-to-noise on faint auroral lines needed to determine electron temperatures and temperature-based ionic and elemental abundances) and do not extend to as large galactocentric radii, we are able to spectroscopically analyse a factor ${\sim}7$ more \HII\ regions. In what follows, we describe how emission line regions are identified and how they are separated into the three main categories, these being \HII\ regions, SNRs, and PNe. 

\subsection{Emission line region identification}
\label{sec:ident}

In a first step, we proceed in identifying and isolating emission line regions across the MUSE mosaic footprint, regardless of their nature. The input for this step is a continuum-subtracted integrated H$\alpha$ map, which reliably traces ionised gas in \HII\ regions and SNRs in the optical regime. 
PNe on the other hand are known to sometimes have very weak Balmer emission \citep{zhang04}, and to identify all of them an additional tracer of highly ionised gas (e.g.\ $[\OIII]\lambda5007$) should be included. We will perform a detailed analysis of the PNe present in the MUSE data in a forthcoming study, and note that the goal of this paper is not that of obtaining a complete PNe census, but rather correctly classifying those PNe that are picked up by the emission line region identification algorithm and excluding them from the \HII\ region and SNR catalogs.

A variety of different methods to identify emission line structures are found in the literature, examples include {\sc HIIPHOT} \citep{thilker00} and Clumpfind \citep{williams94}, the latter having been applied to MUSE data to identify \HII\ regions in NGC~628 \citep{kreckel16}. Here, we proceed in a similar fashion to \citet{dellabruna20}, who used the Python package {\sc astrodendro}\footnote{http://www.dendrograms.org/} to identify bright regions in a MUSE H$\alpha$ map of NGC~7793 ($D \sim 3.4$~Mpc), but we add an extra spectral clustering step to the classical dendrogram hierarchy\footnote{Dendrograms are tree diagrams representing hierarchical structures in e.g.\ astronomical images. Structures are divided into branches (structures that can be further divided into smaller sub-structures) and leaves (the smallest structures that cannot be divided anymore).} of `trunks', `branches', and `leaves' which, as described below, is not ideal at the high spatial resolution achieved with MUSE in NGC~300. 
Hence, to identify emission line regions in the H$\alpha$ map, we use the following simple two-step approach:

\begin{enumerate}
    \item Regions are first broadly isolated by computing a dendrogram of the H$\alpha$ map, exploiting the fact that dendrograms divide the emission in a 2D map into hierarchical structures based on user-defined minimum-flux (with respect to a background) and size thresholds.
    \item The dendrogram structures are then fed into the {\sc scimes} algorithm \citep{colombo15} which groups these into coherent and relevant regions based on a spectral clustering paradigm. 
\end{enumerate}

Dendrograms are heavily dependent on so-called user-defined pruning parameters. On opposite extremes, different pruning parameter choices can lead to either very few large regions encompassing what clearly are individual structures or unrealistically many small fragments. Here, initial pruning parameters are set such that the catalog resulting from the two-step approach approximately contains the expected number of regions in the MUSE footprint based on both visual inspection and on the literature, i.e.\ ${\sim}83$~\HII\ regions \citep{deharveng88}, ${\sim}20$~PNe \citep{stasinska13}, and ${\sim}12$~SNR \citep{millar12}.
The detection flux threshold is set to three times the standard deviation of the H$\alpha$ flux map, i.e.\ about $5\times10^{-18}$ erg~s$^{-1}$~cm$^{-2}$ or about twice the mean H$\alpha$ flux of the map. This comfortably includes the faintest structures identifiable by eye and corresponds to an emission measure of about $60$~pc~cm$^{-6}$, which is somewhat lower than the cutoff value of $82$~pc~cm$^{-6}$ used by \citet{dellabruna20} in NGC~7793 and closer to $50$~pc~cm$^{-6}$ set by \citet{hoopes96} in NGC~300.
We tested different thresholds, finding that higher values do not return the contours of by-eye identified individual regions, whereas lower thresholds lead to large structures encompassing multiple individual regions. The minimum number of pixels is set to~$25$, which, for circular apertures, corresponds to diameters of~${\sim}1\arcsec$, i.e.\ approximately the highest seeing-limited resolution achieved by our observations. With these settings, the resulting dendrogram contains $794$~structures, which are then passed to the clustering algorithm.  

The {\sc scimes} package was originally developed to identify giant molecular cloud structures. At its core is an unsupervised pattern recognition algorithm which, in very general terms, groups together pixels in an image which are considered to be similar to each other. As noted in \citealt{colombo15}, this approach to structure identification overcomes the problem introduced by high spatial resolution observations which causes tools such as dendrograms to overestimate the number of regions (i.e.\ over-divide the input image into too many small structures). We run {\sc scimes} on the dendrogram obtained with the pruning parameters described above, and set the clustering to be performed based on \ha\ flux and to return isolated leaves as well as grouped ones as independent structures. The latter ensures that unresolved emission line regions, e.g.\ PNe, are included in our structure identification.

\begin{figure}
    \centering
    \includegraphics[scale=0.19]{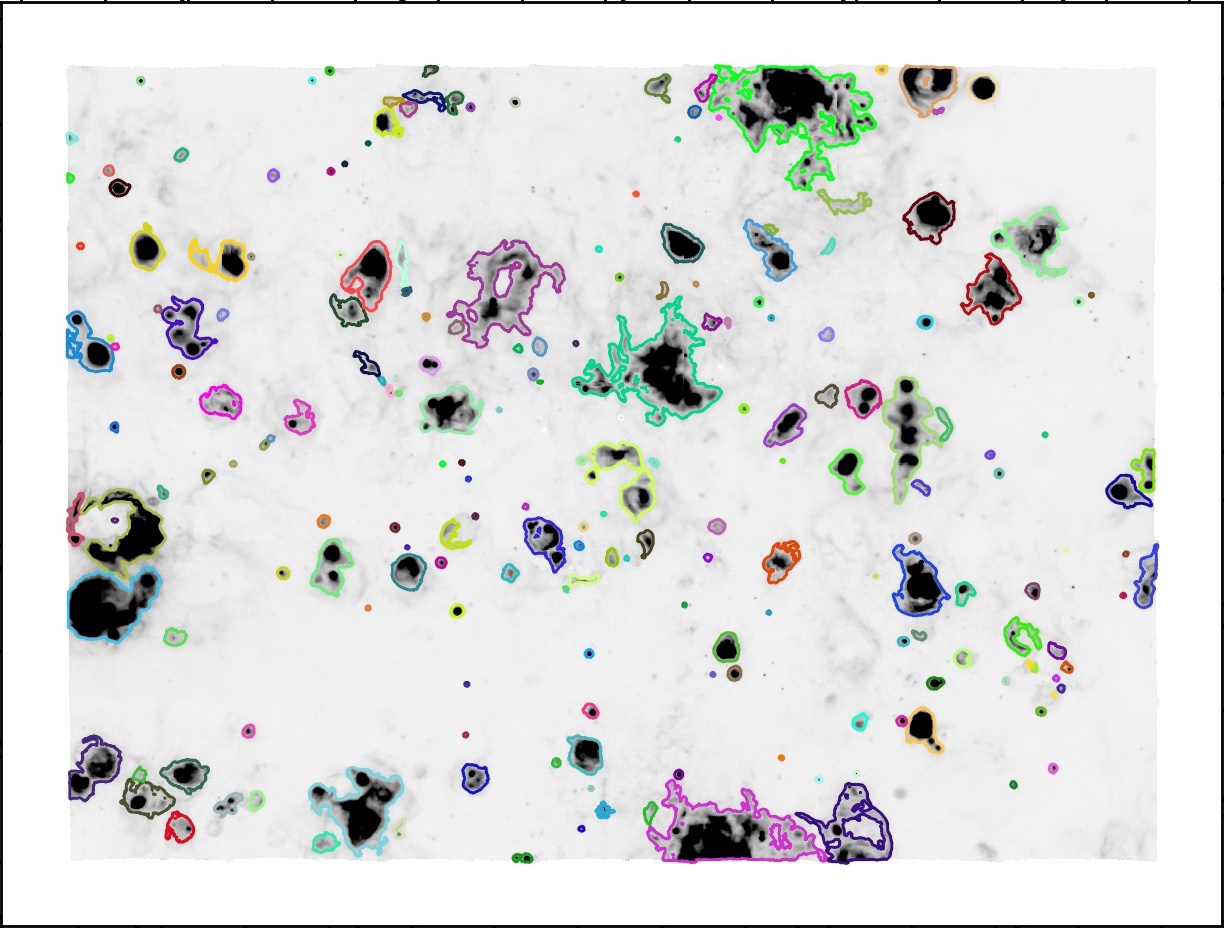}
    \caption{Continuum-subtracted MUSE H$\alpha$ map (scaled from~0 to $2\times10^{-17}$ erg~s$^{-1}$~cm$^{-2}$), the coloured contours correspond to the $204$~emission line regions identified as described in Section~\ref{sec:ident} (colours have no purpose other than facilitating the visual distinction between regions). Emission line region spectra are extracted based on the shown contours.}
    \label{fig:regions}
\end{figure}

With the above described pruning and clustering settings, we obtain an initial catalog of $204$~emission line regions and we extract integrated spectra for all of these based on the identified region contours (see Fig.~\ref{fig:regions}). While some of these regions encompass what are likely multiple regions, these are the minority and only marginally affect the \HII\ region analysis described below, given that our main interest lies in radial trends. As no method is $100$~per~cent accurate in recovering individual structures, in large field-of-view data sets like the one considered here, the vastly greater efficiency of our semi-auto\-mated region identification method is certainly preferred over a subjective by-eye identification. Upon testing different pruning parameters within a sensible range (i.e., as not to return unrealistically many small or just a few large regions), the results from the analyses are unchanged.
From an initial visual inspection of the resulting integrated spectra we eliminate $16$~objects which are either of insufficient quality (this is particularly true for regions in cube L15 which is heavily contaminated by a bright foreground star), or correspond to known stellar sources such as emission line stars \citep[e.g.][]{roth18}, WR stars \citep[e.g.][]{schild03}, as well as PNe \citep[e.g.][]{stasinska13} that have very low signal-to-noise MUSE spectra. The list of eliminated spectra also includes the ultra-luminous \mbox{X-ray} pulsar \mbox{ULX-1} \citep{vasilopoulus19, binder20} as well as a known microquasar \citep{mcleod19,urquhart19}, both of which were picked up in the emission line region identification step due to their interaction with the surrounding ISM (i.e.\ shock-excited gas). The remaining $188$~spectra are passed to the Gaussian fitting routine and subsequent classification scheme described in the next section. 

\subsection{Emission line fitting and region classification}
\label{sec:fitting}

Before fitting the emission lines, the integrated spectra are corrected for Balmer absorption, caused by the unresolved stellar background, using the Python implementation of {\sc pPXF} \citep{cappellari17} together with spectral templates from the MILES library \citep{miles} (as well as a standard list of emission lines to mask). We perform the {\sc pPXF} fit in the range $4750{-}7200$~\AA, thus, excluding redder wavelengths from the fit that contain contaminating residual sky emission. We adopt a MUSE LSF parametrization as in \citet{guerou17} and assume initial guesses of $146$~km~s$^{-1}$ for the systemic velocity (from eq.~8 in \citealt{cappellari17} and a redshift of $z \sim 0.000487$ for NGC~300) and $20$~km~s$^{-1}$ for the velocity dispersion (the latter based on a preliminary inspection of the spectra).
While the main output of {\sc pPXF} fitting consists of the kinematics of the unresolved stellar population, the purpose here is purely to subtract the best-fitting spectral template from the observed integrated emission line spectra and, thus, to obtain a continuum-subtracted and absorption-corrected nebular spectrum for each region. A spatially-resolved study of the stellar kinematics across the entire MUSE mosaic will be presented in a forthcoming publication. We fit the emission lines in each spectrum with the Python package {\sc PySpecKit} \citep{pyspec} assuming single component Gaussians, and correct the obtained line fluxes for extinction using the PyNeb package \citep{luridiana15} based on the Balmer decrement (with an intrinsic \ha/\hb\ ratio of~$2.86$), assuming $R_{V} = 3.1$ and a Galactic extinction curve \citep{cardelli89}. Uncertainties on the ionised gas properties obtained from the Gaussian fits described in the following sections are derived by propagating the errors on the best-fit parameters from {\sc PySpecKit}. As most of the analyses rely on emission line ratios, intrinsic uncertainties are minimised. A further note on the contribution of diffuse ionised gas (DIG) to the integrated spectra. While this paper is not aimed at studying the DIG, it is important to assess whether (and to what extent) the integrated region spectra are contaminated by DIG emission, i.e how much DIG is likely included in the contours we extract spectra from. To this end, we use the region contours (see Fig.~\ref{fig:regions}) to produce a rough DIG map by simply masking all the pixels within the region contours in the continuum-subtracted \ha\ map. We then crudely estimate the amount of DIG in the covered portion of the galaxy by summing the pixel values of the DIG map dividing by the corresponding value of the summed \ha\ map. We obtain a DIG fraction of $\sim$ 47$\pm$2\%, which is in good agreement with \citet[][]{hoopes96} who find a DIG fraction of 53$\pm$5\% for their emission measure threshold of 50 pc cm$^{-6}$. Together with the completeness tests described in Section \ref{sec:hii}, we therefore conclude that the DIG contribution to the extracted region spectra is negligible.

The catalog of $188$~regions obtained as described in Section~\ref{sec:ident} mainly consists of \HII\ regions, SNRs, and PNe. For the \HII\ region and SNR analyses described in the following sections of this paper, the catalog objects therefore need to be classified by type. While several catalogs of \HII\ regions, SNRs, and PNe already exist in the literature, we intentionally do not cross-match our initial emission line region catalog with known sources. This is for two reasons, the first one being that with ongoing large surveys delivering IFU coverage of entire nearby galaxies \citep[e.g.][]{laurie19,emsellem21}, it is in the community's interest to use well-studied galaxies like NGC~300 to find empirical classification methods specifically tailored to the capabilities of the instrument, which can then be used for less well-studied targets. The second reason is that determining the center, size, and boundaries of \HII\ regions at spatial resolutions of roughly $7{-}10$~pc (i.e.\ what is achieved with MUSE in NGC~300) is very subjective and very much dependent on the method that is being used, particularly where regions are in close vicinity or even overlap, meaning that \HII\ region catalogs of the same galaxy but from different papers can vary significantly. This is particularly noticeable for \HII\ region catalogs that have been defined by eye versus those that result from more sophisticated approaches relying on structure identification algorithms. We do perform a literature cross-match for SNRs and PNe in a second step, which serves the purpose of validating our classification method. 

\begin{figure}
    \centering
    \includegraphics[scale=0.47]{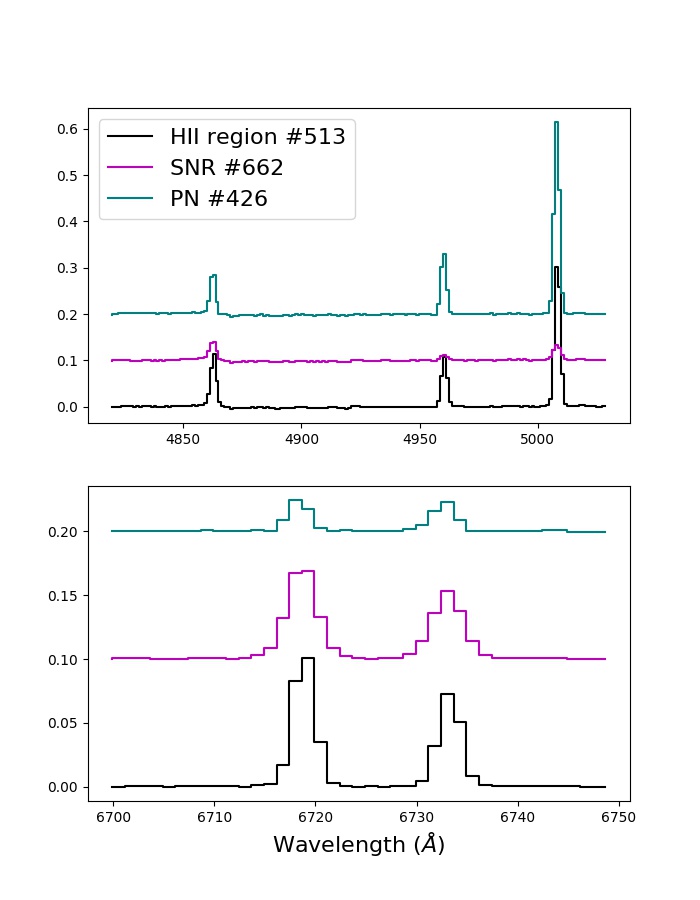}
    \caption{Normalised nebular spectra of an \HII\ region (black), a SNR (magenta), and a PN (blue), classified as described in Section~\ref{sec:fitting}. Flux units are arbitrary. The spectra are cropped to relevant wavelength ranges, i.e.\ covering the H$\beta$ and $[\OIII]\lambda4959,5007$ lines (top), and the $[\SII]\lambda6717,31$ lines (bottom). Details about the three objects are given in the Appendix tables.}
    \label{fig:spectra}
\end{figure}

In what follows, we describe the classification scheme used to disentangle between three main groups of objects, i.e.\ \HII\ regions, SNRs, and PNe. For illustration purposes, Fig.~\ref{fig:spectra} shows normalised spectra representative of an \HII\ region, a SNR, and a PN classified as per the below. The spectra in this figure are cropped to wavelength ranges covering relevant emission lines, i.e.\ the H$\beta$ and $[\OIII]\lambda4959,5007$ lines (top panel), and the $[\SII]\lambda6717,31$ lines (bottom panel). Details of the three objects (coordinates, identifiers, emission line ratios, etc) are given in Appendix~\ref{app:tables}.

\subsubsection{Supernova remnants}
\label{sec:snr}

The most commonly used line ratio diagnostic to identify SNRs in external galaxies is $[\SII]/\ha$, as it gives a relative measure of the ionisation stages of sulphur within a given region. In \HII\ regions, where the gas is mostly photoionised, a larger fraction of sulphur is expected to be in the higher excitation state (S$^{++}$), and $[\SII]/\ha$ ratios are typically around $0.1$ \citep{long18}. In SNRs, where shocks contribute to populating S$^{+}$, $[\SII]/\ha$ values are expected to be $0.4$ or higher \citep{allen08}. \citet{blair97} used the $[\SII]/\ha$ diagnostic to compile a list of $28$~SNR candidates in NCG~300, which \citet{millar12} later reduced to $22$~objects based on additional line ratios and optical data. Twelve of the \citeauthor{millar12} SNRs are within the MUSE mosaic, although this list includes what is now known to be a microquasar (their source~S10; see \citealt{mcleod19,urquhart19}), leaving eleven reasonable SNR candidates within the MUSE footprint. In addition to the $[\SII]/\ha > 0.4$ criterion (which alone is not sufficient to separate SNRs from \HII\ regions, as the latter can have values well above the $0.4$ threshold; see \citealt{long18}), we also inspect [\SII] line widths. At the spectral resolution of MUSE (${\sim}120$~km~s$^{-1}$ in the H$\alpha$ regime), \HII\ region line widths \citep[typically ${<}40$~km~s$^{-1}$;][]{kewley19} are very likely unresolved (i.e.\ instrumental), while the Doppler broadened lines of ${>}100$~km~s$^{-1}$, characteristic of radiative shocks in younger SNRs \citep[e.g.][]{long18}, are expected to be resolved in our observations. Older SNRs with line widths ${<}100$~km~s$^{-1}$ might therefore be misclassified, however, these also are typically very faint and would likely fall below our detection threshold. 

\begin{figure}
    \centering
    \includegraphics[scale=0.45]{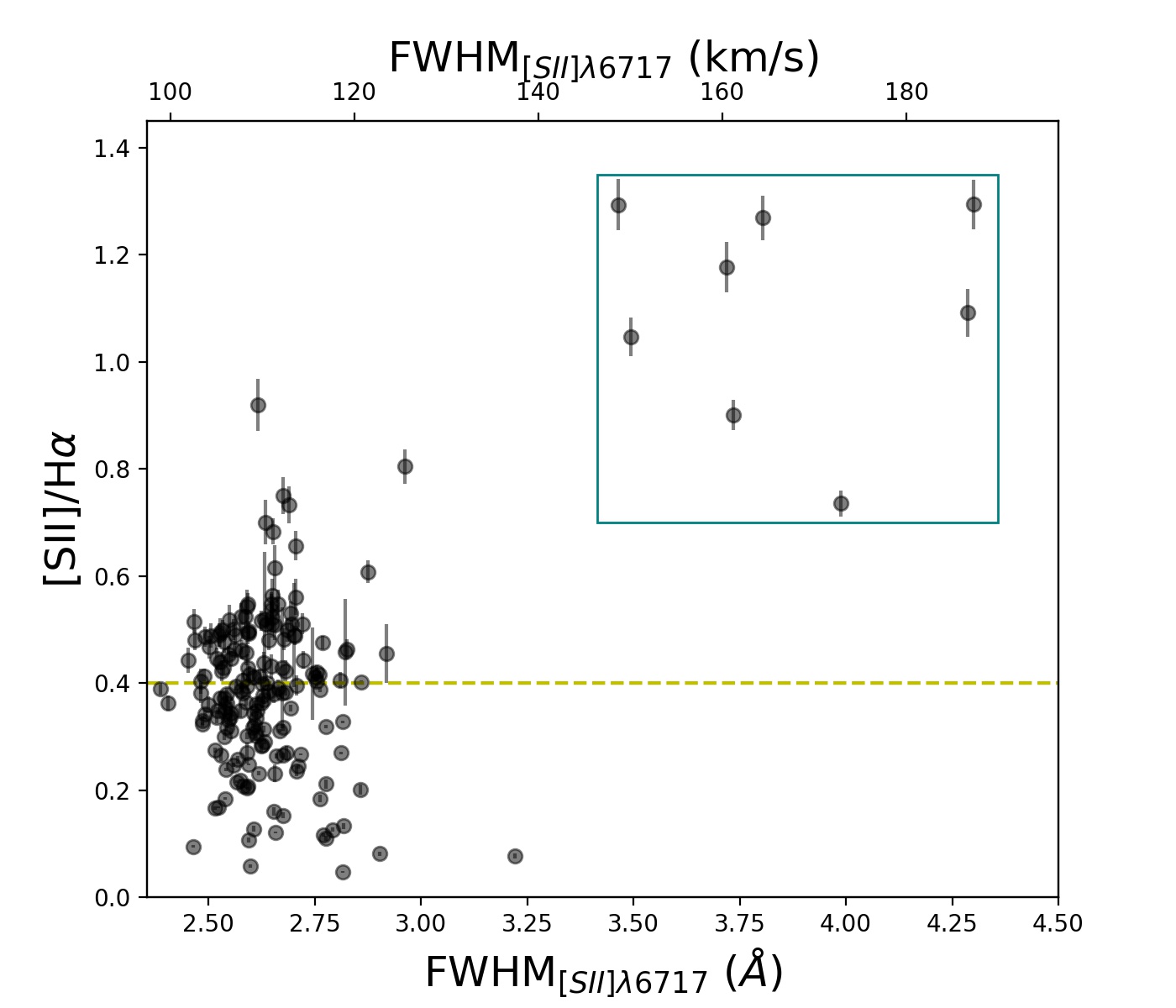}
    \caption{SNR candidate selection. The $[\SII]/\ha$ emission line ratio as a function of the FWHM of the $[\SII]\lambda6717$ line as obtained from the Gaussian fitting routine. The yellow dashed line indicates the traditional criterion to identify SNR candidates, $[\SII]/\ha > 0.4$. The box (without a quantitative purpose) highlights the region of parameter space occupied by SNRs in NGC~300, see text in Section~\ref{sec:snr}.}
    \label{fig:s2ha}
\end{figure}

As is shown in Fig.~\ref{fig:s2ha}, the parameter space spanned by the full width half maximum (FWHM) of the $[\SII]\lambda6717$ line, $\mathrm{FWHM}_{[\SII]}$, and the $[\SII]/\ha$ ratio clearly separate the detected emission line regions into two distinct populations. In this parameter space, SNRs are expected to reside in the upper right quadrant due to their intrinsically broader lines and higher $[\SII]/\ha$ ratios. To confirm the classification of regions identified as SNRs in the $\mathrm{FWHM}_{[\SII]} {-} [\SII]/\ha$ parameter space, indicated by the (arbitrary) blue box in Fig.~\ref{fig:s2ha}, we cross-match their coordinates with the eleven \citet{millar12} sources that lie within the MUSE field, and find that 7 out of the 8 sources identified here indeed correspond to previously known SNRs. The unmatched source (id~\#538, see Table~\ref{tab:snr}) is the only one of the 8 to be unresolved in our data, and we therefore exclude it from further analyses. We do however assign this a SNR candidate flag in our catalog. Thus, 7 of the 11 \citeauthor{millar12} sources in our mosaic are correctly classified in the $\mathrm{FWHM}_{[\SII]} {-} [\SII]/\ha$ parameter space. For NGC~300 we therefore define SNRs as those objects satisfying both $\mathrm{FWHM}_{[\SII]} > 3.4$~\AA\ and $[\SII]/\ha\ > 0.4$, where the ${>}3.4$~\AA\ is empirical and corresponds to a velocity line width of ${\sim}150$~km~s$^{-1}$, reasonable for SNRs and not expected of \HII\ regions.

Of the remaining 4 \citeauthor{millar12} sources that we do not recover (S08, S09, S19, and S22), S08 and S09 do not show line broadening and are consistent with being \HII\ regions in our sample (based on their line ratios, line widths, and upon visual inspection), S22 is a faint structure that lies below our detection threshold (and upon further inspection does not show line ratios and line widths consistent with a SNR), and S19 is grouped into a contour with an adjacent \HII\ region (we will discuss the implication of this in Section~\ref{sec:hii}).

The above method to distinguish SNR from other emission line regions is thus robust, but we recommend that, if the purpose is that of identifying unknown SNRs rather than removing them from an \HII\ region sample, the detection threshold in the dendrogram emission line region identification step is lowered to also include potentially low-brightness SNRs. We also recommend using a [\SII] map in addition to the H$\alpha$ map when specifically identifying SNRs. 

\subsubsection{Planetary nebulae}
\label{sec:pne}

With the SNRs being removed from the emission line catalog according to the empirical separation described in Section~\ref{sec:snr}, we now seek to disentangle \HII\ regions from PNe. Countless pairs of line ratios and other empirical relations have been used in the literature, e.g.\ the traditional BPT diagnostic plot \citep{baldwin81}, the $\ha/[\SII]$ versus $\ha/[\NII]$ diagnostic \citep{riesgo06}, or more recently, with MUSE data of NGC~628, empirical narrow-band criteria \citep{kreckel17}. 

\begin{figure}
    \centering
    \includegraphics[scale=0.48]{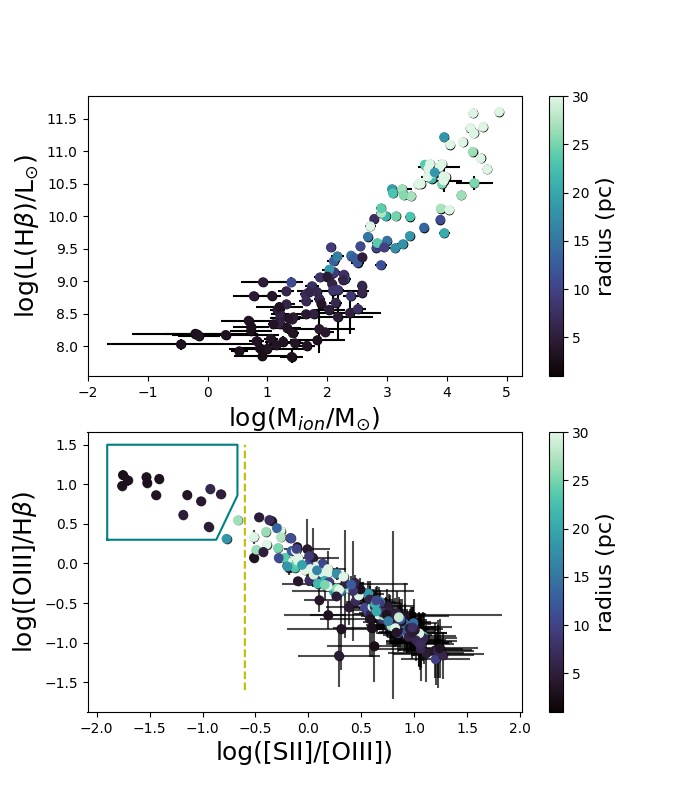}
    \caption{{\it Top panel}. $L(\hb)$ as a function of the ionised gas mass, $M_\mathrm{ion}$, colour-coded by region radius. {\it Bottom panel}. $[\OIII]/\hb$ as a function of $[\SII]/[\OIII]$. The yellow dashed line (at $\log([\SII]/[\OIII]) = -0.6$) empirically separates NGC~300 PNe from \HII\ regions when combined with an additional region size requirement (colour scale). The teal box encompasses the sources classified as PNe, see text in Section~\ref{sec:pne}.}   
    \label{fig:s2o3}
\end{figure}

Here, we exploit the fact that PNe typically exhibit large $[\OIII]/\hb$ ratios \citep[e.g.][]{baldwin81} and, due to their hot central stars, large degrees of ionisation, together with the fact that at the distance of NGC~300 PNe are spatially unresolved and, therefore, appear as point sources (with PNe typically having radii ${\lesssim}1$~pc, \citealt{jacob13}). 
This is shown in the lower panel of Fig.~\ref{fig:s2o3}, where we use the $[\SII]/[\OIII]$ ratio as a proxy for $[\OII]/[\OIII]$, the degree of ionisation tracer (given that our observations are not deep enough to detect the $[\OII]\lambda7320,7330$ lines with sufficient signal-to-noise). Here, PNe reside in the upper left quadrant due to their high degrees of ionisation and their enhanced $[\OIII]/\hb$ ratios, and are less confused with compact (i.e.\ unresolved) \HII\ regions than in the upper panel. We therefore define empirical separation criteria and select PNe candidates as those objects with $\log([\SII]/[\OIII]) \lesssim -0.6$, $\log([\OIII]/\hb) \gtrsim 0.2$, and having radii ${<}7$~pc (i.e.\ being unresolved). This selects the sources residing in the teal box in Fig.~\ref{fig:s2o3}.

Several more pairs of quantities to disentangle \HII\ regions from PNe were explored in NGC~300 by \citet{stasinska13}, who found a clear sequence in the H$\beta$ line luminosity, $L(\hb)$, versus the ionised gas mass, $M_\mathrm{ion}$, ranging from PNe to compact \HII\ regions and to evolved \HII\ regions. While we clearly recover this sequence, shown in the upper panel of Fig.~\ref{fig:s2o3} (data points are colour-coded by region radius), the large uncertainties in estimating the electron density, $n_\mathrm{e}$, and the rather coarse assumptions on stellar initial mass function (IMF) sampling when converting the H$\alpha$ luminosity to an ionising flux, $Q_{L(\ha)}$ (see Section~\ref{sec:press}), do not allow reliable estimates of $M_\mathrm{ion}$, which we compute as in \citetalias{mcleod20},
\begin{equation}\label{eq:mion}
    M_\mathrm{ion} \approx Q_{L(\ha)} m_\mathrm{p} / (n_\mathrm{e} \alpha_\mathrm{B})
\end{equation}
\noindent with the proton mass $m_\mathrm{p}$ and the Case~B coefficient $\alpha_\mathrm{B}$ (corresponding to optically thick nebulae), and assuming negligible dust absorption. Electron densities, $n_\mathrm{e}$, are computed via {\sc pyneb} from the $[\SII]\lambda6717 / [\SII]\lambda6731$ line ratio assuming a temperature of $10^{4}$~K\footnote{This assumption results in the derived electron densities and ionised gas masses being lower and upper limits, respectively, given that PNe in NGC~300 can have electron temperatures up to a factor ${\sim}1.4$ higher \citep[][]{stasinska13}.}, and while \HII\ regions in our sample have measured electron density uncertainties, $\sigma_{n_\mathrm{e}}$, on the order of ${\sim}20$~per~cent, the PNe have $\sigma_{n_\mathrm{e}}$ ranging from about 25~per~cent to over 200~per~cent. Given the consequently large uncertainties on $M_\mathrm{ion}$ and given that there is no clear separation between PNe and \HII\ regions, we do not use this parameter pair to disentangle the two types of regions. 

To confirm the validity of our PNe selection criteria, we cross-match the resulting list of 13~PNe candidates with the 18~PNe from the \citet{stasinska13} catalog falling within the MUSE field-of-view, confirming all but one of our PNe candidates, i.e.\ 12 out of the expected 18~PNe from \citeauthor{stasinska13} are correctly identified. The unconfirmed source of our 13~candidates (id~\#156, see Table~\ref{tab:pne}) which is not in the \citeauthor{stasinska13} catalog is consistent with being a PN based on its line ratios which place it well above the extreme starburst lines in the BPT diagram (Fig.~\ref{fig:bpt}), and we therefore assign it a PN flag. This leaves 6 of the \citeauthor{stasinska13} PNe that we do not recover with our selection. Of these, 4 are below the detection threshold and are therefore not recovered in the dendrogram, one falls within an \HII\ region contour, and the last is consistent with being an \HII\ region based on its emission line ratios and its radius of ${\sim}11$~pc (which is therefore well resolved in the observations unlike the other PNe). 

This shows that our PNe selection method is very robust, considering that with the source detection threshold and other dendrogram pruning parameters we recover all of the known PNe in the field that were picked up in the emission line region identification step. Again, we note that the goal of this study is not to compile a complete census of PNe, but rather to remove the ones that fall within our detection algorithm from the \HII\ region sample. By lowering the detection threshold in the dendrogram analysis and by including the [\OIII] map to identify regions, the four undetected PNe would likely have been correctly matched. Conversely, this implies that we also do not detect other fainter \HII\ regions, which however does not further impact our analysis given their expected low signal-to-noise ratios and thus high uncertainties.

\subsubsection{\texorpdfstring{\HII}{HII} regions}
\label{sec:hii}

\begin{figure}
    \centering
    \includegraphics[scale=0.43]{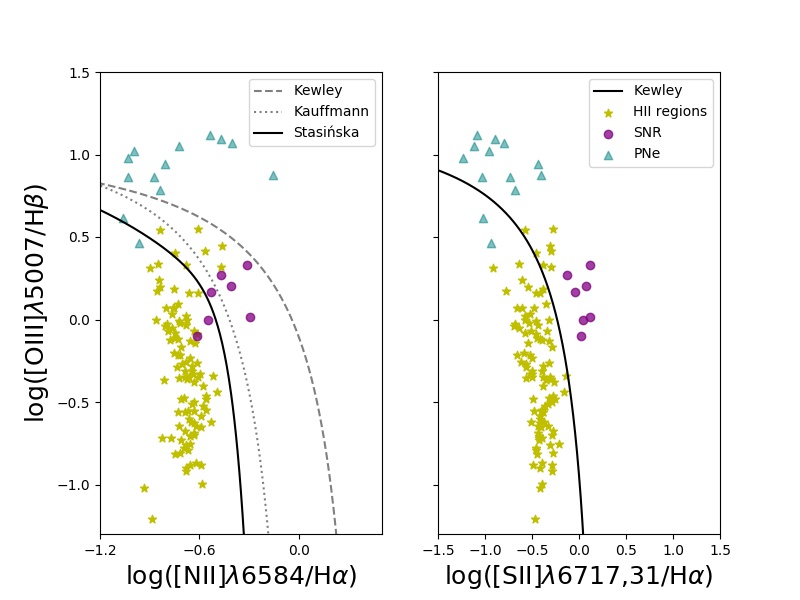}
    \caption{BPT diagrams of the emission line regions identified in this work: \HII\ regions (yellow stars), PNe (teal triangles), and SNR (purple circles) as classified according to Section~\ref{sec:regions}. Classical lines separating AGN-dominated and star formation-dominated regimes from the literature are shown \citep[][]{kewley01,kauffmann03,stasinska06}.}    
    \label{fig:bpt}
\end{figure}

With the SNR and PNe selection criteria described above, the initial emission line region catalog of $188$~objects is reduced to $103$~spatially resolved sources (i.e.\ after removing SNRs and PNe we place an additional constraint on the size of the regions by requiring a radius $r > 7$~pc to only include spatially resolved sources, bringing the number down to $103$~regions), which are therefore classed as bona fide \HII\ regions and used for the subsequent analyses. Their line ratios are consistent with and place them in the expected BPT diagram space occupied by \HII\ regions. This is illustrated in Fig.~\ref{fig:bpt}, where, in addition to the extreme starburst lines from \citet{kewley01} and \citet{kauffmann03} that are widely used to separate star formation- from AGN-dominated galaxies, we also show the separation line proposed by \citet{stasinska06} for local star-forming galaxies. 

\begin{figure}
    \centering
    \includegraphics[scale=0.4]{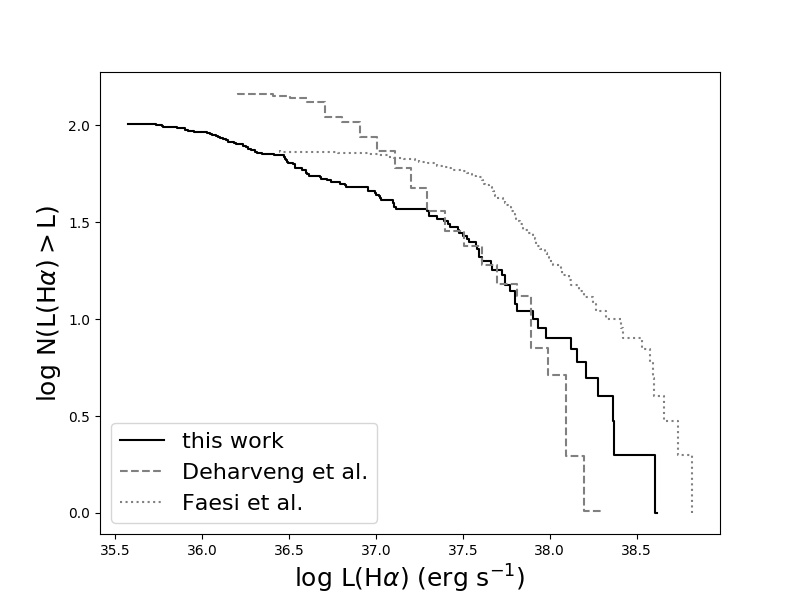}
    \caption{Cumulative \ha\ luminosity function of the \HII\ regions as identified in this work (black solid line), in \citet{deharveng88} (grey dashed line), and from table~2 of \citet{faesi14} (grey dotted line).}
    \label{fig:lumfunc}
\end{figure}

As mentioned above, cross-matching the resulting \HII\ region catalog with catalogs from the literature is not as straightforward as for PNe and SNRs. While a spatial comparison with the location of the ${\sim}83$~\HII\ regions from \citet{deharveng88} that fall within the MUSE field-of-view shows a qualitative good agreement, it is clear that the dendrogram+clustering structure identification recovers smaller and more compact regions that do not appear in the \citeauthor{deharveng88} catalog on the one hand, but on the other hand tends to group together a handful of regions into larger complexes. A more quantitative measure for the robustness of the identification approach in recovering NGC~300 \HII\ regions is to compare population statistics with previous studies, e.g.\ the \HII\ region H$\alpha$ luminosity function, as is shown in Fig.~\ref{fig:lumfunc}. This shows good agreement with \citet{deharveng88}, in particular in the $10^{37}~\textrm{erg~s}^{-1} < L(\ha) < 10^{38}~\textrm{erg~s}^{-1}$ regime. At the high luminosity end, the dendrogram approach used here leads to a slight horizontal shift towards higher luminosities due to the grouping of some regions, while at the low luminosity end, the much higher spatial resolution of the MUSE data (compared to \citeauthor{deharveng88}) leads to a flatter tail. Also shown in Fig.~\ref{fig:lumfunc} is the \HII\ region \ha\ luminosity function compiled from values given in table~2 of \citet{faesi14}, which is systematically shifted towards higher luminosities, likely due to the fact that \citeauthor{faesi14} use fixed $13\farcs5$ apertures (${\sim}130$~pc at $D = 2$~Mpc) for most \HII\ regions, therefore, often overestimating the \ha\ flux.

\begin{figure}
    \centering
    \includegraphics[scale=0.45]{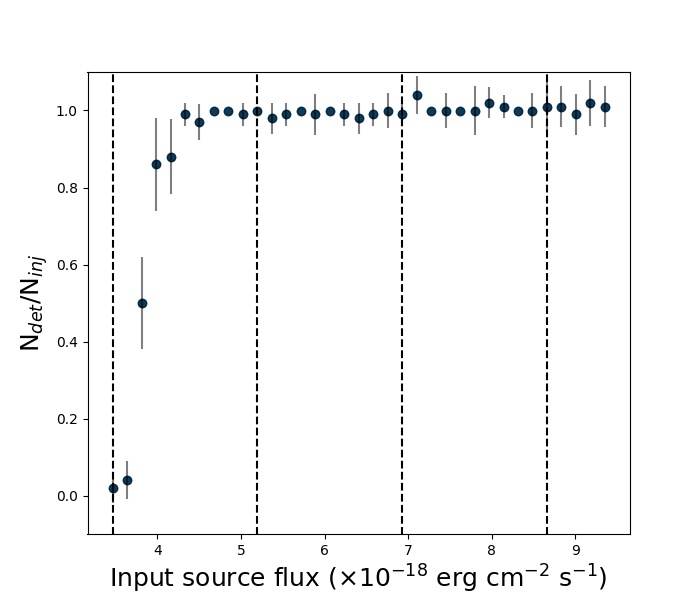}
    \caption{The detection fraction (number of detected Gaussian sources, N$_{det}$ over the number of injected Gaussian sources, N$_{inj}$), as a function of input source flux. Vertical dashed lines represent 2$\sigma$, 3$\sigma$, 4$\sigma$, and 5$\sigma$ fluxes, with $\sigma = 1.7\times10^{-18}$ erg cm$^{-2}$ s$^{-1}$ being the mean noise level of the \ha\ map. The catalog is $>$99\% complete for source fluxes $>2.5\sigma$.} 
    \label{fig:complete}
\end{figure}

Based on the above, the \HII\ region catalog is likely complete in the regime of small region radii down to the spatial resolution limit of about 7 pc. However, we are likely incomplete in the regime of large region radii, due to the fact that larger and brighter regions are less easily detected than smaller, fainter regions, i.e., for given fluxes (respectively, for given radii), regions with increasing radii (respectively, decreasing fluxes) will eventually fall below the surface brightness cut. To further investigate the completeness, we proceed in a similar fashion to \citet[][]{rosolowsky2010} and produce a simulated \ha\ map, inject Gaussian sources, and run our detection algorithm to assess the source recovery. While Gaussian sources are not ideally matched to the flux distributions observed in the real \ha\ map (in which \HII\ regions have a variety of different morphologies, from roughly circular with a central peak to open and filamentary), this remains a worthwhile test to do to assess completeness. The simulated map has the same dimension (i.e., number of pixels, WCS, pixel size, etc.) and noise properties as the observed image used for the structure identification. We then run the region identification algorithm on the synthetic map by injecting Gaussian sources (with radii evenly distributed between the observed values) with increasing surface brightnesses into a non-crowded FOV, to assess the impact of low surface brightness regions.  We do not vary pruning parameters (minimum number of pixels for independent structures and the leaf significance) in this test, as these are not driving the (in-)completeness at large region sizes.  To estimate the degree of completeness we compare the number of injected to the number of detected sources as a function of input parameters from 10 bootstrap iterations to estimate uncertainties on the detection fraction. The completeness fraction as a function of input Gaussian source flux is shown in Fig.~\ref{fig:complete}, indicating that the catalog is $>$99\% complete for source fluxes $>$2.5$\sigma$, with $\sigma = 1.7\times10^{-18}$ erg cm$^{-2}$ s$^{-1}$ being the mean noise level of the \ha\ map. This means that with the pruning parameters set as described in Section \ref{sec:ident}, we recover most sources brighter than $>$2.5$\sigma$. 
In terms of purity, we note that contamination from PNe to the resulting \HII\ region catalog is unlikely, as PN contaminants would have been removed with the size constraint. SNR contamination is more likely, given that very old SNRs would have velocities below the MUSE resolution, as well as low [SII]/\ha\ ratios.

We caution that while our emission line region classification method (i.e., the tetrad of [SII]/\ha, [OIII]/\hb, [SII]/[OIII], and $\mathrm{FWHM}_{[\SII]}$) can be applied to other nearby galaxies, the numerical values for thresholds and criteria (i.e.\ the dashed yellow separation lines in Figures~\ref{fig:s2ha} and~\ref{fig:s2o3}) are NGC~300 specific and are likely different in different environments as the strength of emission lines is affected by factors such as metallicity and the resolved stellar population within the regions. To make the presented classification scheme widely applicable, it would need to be augmented with photoionisation models, and tested on spectroscopic data sets of nearby galaxies that differ not only in terms of galaxy properties (e.g.\ mass, metallicity, type, etc.) but also in terms of distance which affects resolution.

\section{Feedback-driven gas in \texorpdfstring{\HII}{HII} regions}
\label{sec:gas}

Having compiled a catalog of \HII\ regions, we now exploit the full range of nebular emission lines covered by the MUSE observations to characterise the feedback-driven gas in the regions. With the aim of exploring the existence of a relation between \HII\ region properties and the impact of massive feedback-driving stars that have formed within them, of particular interest are gas-phase abundances and ionisation properties. For this, in Section~\ref{sec:abun} we first derive key gas properties, and then link these to feedback-related pressure terms within the regions in Section~\ref{sec:press}.

\subsection{Gas-phase abundances and ionisation properties}
\label{sec:abun}

To derive gas-phase abundances in \HII\ regions, the direct temp\-era\-ture-based method is generally preferred \citep[e.g.][]{peimbert17}. It has been used to derive abundances in NGC~300 by \citet[][28 \HII\ regions]{bresolin09}, \citet[][9~compact \HII\ regions]{stasinska13}, and more recently by \citet[][7~\HII\ regions]{toribio16}, confirming that NGC~300 has a negative metallicity gradient, as discussed in \citet{deharveng88}. However, this method relies on auroral line measurements with sufficient signal-to-noise to determine electron temperatures, and while the MUSE wavelength range includes some of these lines (e.g.\ $[\NII]\lambda5754$, $[\SIII]\lambda6312$), they are typically faint and get fainter towards the higher metallicity regime \citep{curti17} of the central regions observed in NGC~300 (which range between about half solar metallicity, similar to the Large Magellanic Cloud (LMC), and ${\sim}0.6$~solar; \citealt{bresolin02}). The MUSE observations used in this work are not sufficiently deep to obtain reliable auroral line detections, and we therefore use the so-called strong line method to determine abundances. This method relies on calibrations obtained either from theoretical calculations or by fitting observed relations between different strong line ratios and abundances derived from the direct method. 

There is a rich history of both auroral and theoretical calibrations in the literature for a variety of different metal\-licity-sensi\-tive strong line ratios, and a review of these is given in \citet{kewley19}. Here, we use the N2~ratio ($\equiv [\NII]\lambda6584/\ha$) which, among the possible strong line ratios covered by MUSE, is the least sensitive to reddening and flux calibration issues due to the vicinity of the two lines. More importantly, it makes use of only two emission lines, therefore minimising possible dependencies on other emission lines of correlations discussed below.
Another possible strong line ratio available from the MUSE coverage is O3N2 ($\equiv ([\OIII]\lambda5007/\hb) / ([\NII]\lambda6584/\ha)$), which is valid across a larger metallicity regime than~N2, but the line ratio itself suffers from a strong dependence on the ionisation parameter which would propagate to the derived metallicities. The ionisation parameter dependence of the N2~ratio is lower, with the N2-derived metallicity varying by about 1 order of magnitude with the ionisation parameter \citep{kewley19}. Composite diagnostic strong line ratios have been proposed to overcome the ionisation parameter dependence (e.g.\ \citealt{dopita16}, which combines the N2 and $[\NII]/[\SII]$ ratios), as well as theoretical calibrations which include an ionisation parameter correction (e.g.\ \citealt{kewley19}). The former introduces a non-negligible amount of scatter when compared to abundances derived from N2 only, and we therefore do not use it. Instead, we use the empirical N2 calibration given in \citet{marino13}, but explore the difference between this and the ionisation-corrected \citeauthor{kewley19} calibration in Appendix~\ref{app:abun}. 

Other MUSE studies of nearby galaxies \citep[e.g.][]{kreckel19} computed oxygen abundances using the empirical calibrations given in \citealt{pilyugin16}, which combine three strong line ratios in their calibrations. For example, the \citeauthor{pilyugin16} $S$~calibration is based on a combination of $[\SII]/\hb$, $[\OIII]/\hb$, and $[\NII]/\hb$. Because most \HII\ region properties computed in this paper are based on strong lines, we refrain from using metallicity calibrations that use many lines in order to avoid introducing systematic dependencies when analysing relations (e.g.\ the pressure of the ionised gas, $P_{\HII}$, is proportional to the $[\SII]$ lines, see Section~\ref{sec:press}). 

\begin{figure}
    \centering
    \includegraphics[scale=0.43]{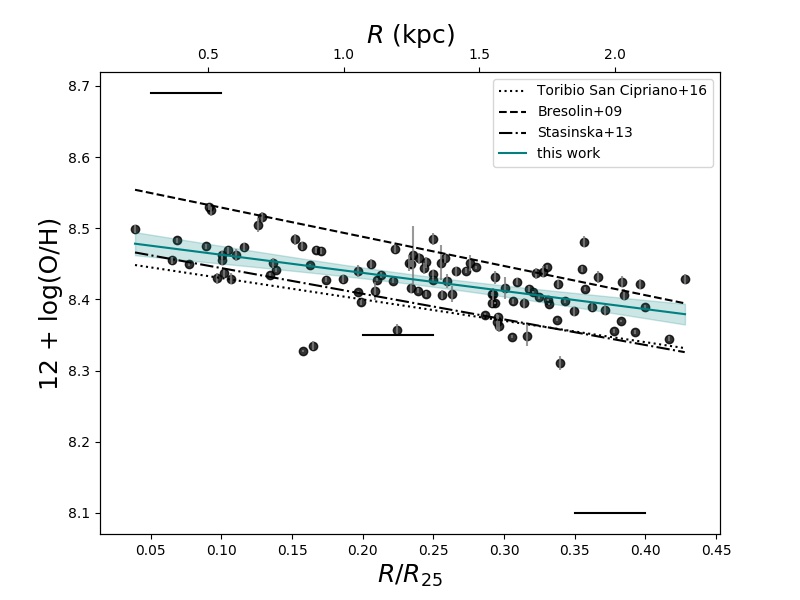}
    \caption{Metallicity gradient of NGC~300 as measured from \HII\ regions, see Section~\ref{sec:abun}. The teal line is a linear fit to the data ($\mathrm{12+log(O/H)} = 8.50\,(\pm0.01) - 0.25\,(\pm0.04) \, R/R_{25}$), the 95~per~cent confidence region is shaded. The dotted, dashed, and dot-dashed lines correspond to the abundance gradients as determined from the direct method by \citet{toribio16}, \citet{bresolin09}, and \citealt{stasinska13}, respectively. The three solid black lines mark solar \citep[${\sim}8.69$;][]{asplund05}, LMC (${\sim}8.35$), and SMC (${\sim}8.10$) metallicities.}    
    \label{fig:metgrad}
\end{figure}

Fig.~\ref{fig:metgrad} shows the abundance gradient (derived from the N2~ratio) as traced by the \HII\ regions, which, in the central part of NGC~300 covered by the MUSE mosaic, have abundances ranging from ${\sim}2/3$ solar to ${\sim}1/2$ solar. We perform a linear fit, obtaining $\mathrm{12+log(O/H)} = 8.50\,(\pm0.01) - 0.30\,(\pm0.03) \, R/R_{25}$. The slope of $-0.25\,(\pm0.04)$ is, within errors, in good agreement with that of $-0.30\,(\pm0.08)$ found by \citet{toribio16} and with the slope of $-0.36\,(\pm0.05)$ reported in \citet{stasinska13}, while the slope of $-0.41\,(\pm0.03$) reported in \citet{bresolin09} is slightly steeper (all three of these studies derive \HII\ region abundances from the direct method). The intercept of $8.50\,(\pm0.01)$ is also in good agreement (within errors) with those found by \citet{toribio16} and \citet{stasinska13}, and about $0.15$~dex lower than that of \citet{bresolin09}. We note that differences in slope and intercept with the literature are likely due to two key factors. Firstly, we use the strong line method to derive abundances, while the values we compare our results to have been derived using the direct method. As discussed in \citet{bresolin09}, different strong line ratios can sometimes lead to drastically different intercepts and slopes, with N2 the one showing the best agreement with the abundance gradient from the direct method. Indeed, the calibration of the N2~ratio is based on the $[\OIII]\lambda4363$ line which, together with $[\OIII]\lambda5007$, is temp\-era\-ture-sensi\-tive. Secondly, the MUSE data only cover the inner part of NGC~300 (for reference, the \citeauthor{bresolin09} study extends to about $1R_{25}$), and additional MUSE data covering the outer parts of the disc would be needed to better constrain the abundance gradient. 

\begin{figure}
    \centering
    \includegraphics[scale=0.55]{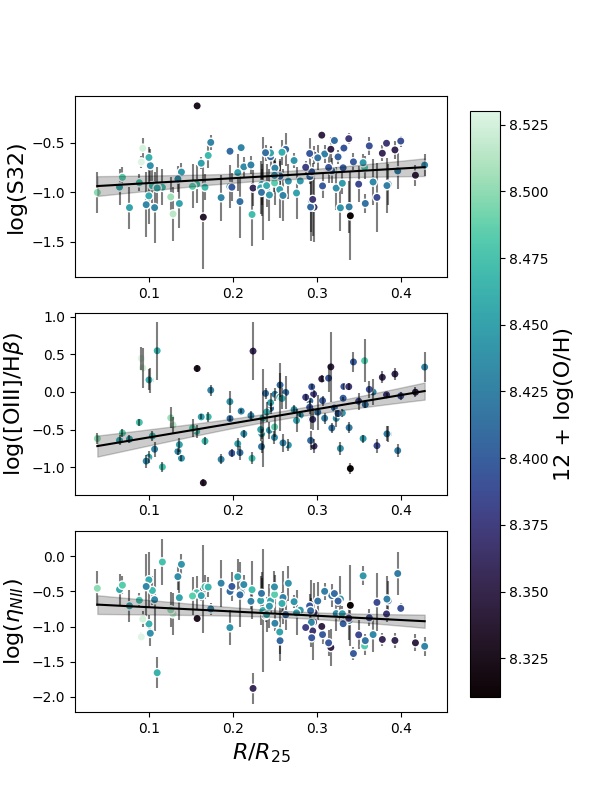}
    \caption{\HII\ region ionised gas properties as a function of galactocentric radius and colour-coded by the gas-phase abundance abundance. S32 and $[\OIII]/\hb$ (top and middle panels) trace ionised and highly ionised gas, respectively, while the softness parameter, $\eta_{\NII}$ is inversely correlated with the effective temperature of the excitation sources. Solid black lines are linear fits to the data, shaded areas are 95~per~cent confidence regions.}
    \label{fig:gradients}
\end{figure}

Another key property that can be determined using optical emission line ratios is the degree of ionisation of the gas within the \HII\ regions. With the observed metallicity gradient, this is indeed an interesting quantity to explore, as lower metallicities imply that stars are hotter and have higher photon fluxes, which therefore translate into higher degrees of ionisation. Here, we use the ratio of the two sulphur ionisation states within the MUSE wavelength coverage, $\mathrm{S32} \equiv [\SIII]/[\SII]$, which is considered a good tracer of the degree of ionisation since the actual line ratio itself is significantly less dependent on metallicity than, e.g.\ $[\OIII]/[\OII]$ \citep{kewley19}, particularly in the metallicity range observed here ($8.3 \lesssim \mathrm{12+log(O/H)} \lesssim 8.6$). The top panel of Fig.~\ref{fig:gradients} shows the degree of ionisation, traced by S32, as a function of galactocentric radius. The linear fit (black line) yields a positive gradient of $+0.69\,(\pm0.19)$, showing that the degree of ionisation increases towards the outer parts of the MUSE field-of-view (i.e.\ towards $0.45 R_{25}$).


We further explore the ionisation properties of \HII\ regions within the covered portion of NGC~300 in terms of the radiation hardness. Similar to \citet{perez-montero09} and based on \citet{vilchezpagel88}, we use a `ratio of ratios' to evaluate the radiation softness parameter, 
\begin{equation}\label{eq:soft}
    \eta_{\NII} = \frac{\mathrm{N}^{+}/\mathrm{O}^{2+}}{\mathrm{S}^{+}/\mathrm{S}^{2+}} \sim  \frac{[\NII]\lambda6584/[\OIII]\lambda5007}{[\SII]\lambda6717,31/[\SIII]\lambda9068}~.
\end{equation}
Here we have used [\NII] as a proxy for [\OII], which is used by \citet{perez-montero09} but is not covered with sufficient signal-to-noise by our observations, and we therefore distinguish the modified softness parameter as defined by Eq.~\ref{eq:soft} from the one used by \citeauthor{perez-montero09} by adding the subscript~\NII. The softness parameter, $\eta$ does not dependent on local effects of the ionisation parameter and density, it is sensitive to the effective temperature of the stars ($\eta \sim 1/T_\mathrm{eff}$) within a given region, considering the difference of the ionising potentials involved in the ratio (i.e.\
$E_\mathrm{ion}\,(\mathrm{S}^{+}) \sim 10.4$~eV, $E_\mathrm{ion}\,(\mathrm{S}^{2+}) \sim 23.3$~eV, $E_\mathrm{ion}\,(\mathrm{N}^{+}) \sim 14.5$~eV, $E_\mathrm{ion}\,(\mathrm{O}^{2+}) \sim 35.1$~eV),
and its value decreases with increasing hardness of the radiation field. \citeauthor{perez-montero09} find that individual nearby galaxies have negative $\eta$ gradients with varying slopes.

\citet{bresolin09} find a metallicity trend in the hardness of the ionising radiation in NGC~300. While a quantitative comparison with the linear least-square fit found by these authors is not possible due to the modification of the softness parameter used in this work, a qualitative comparison confirms the trend of $\eta_{\NII}$ with metallicity (Fig.~\ref{fig:gradients}, lower panel). Together with the trend found for S32, this indicates that stars have higher effective temperatures in the outer \HII\ regions of NGC~300 probed by the MUSE data, producing harder ionising photons and resulting in higher degrees of ionisation. 

Fig.~\ref{fig:gradients} shows an overall picture of the radial and metallicity trends of the ionisation state and radiation hardness in the \HII\ regions, and we additionally show the $[\OIII]/\hb$ ratio as another tracer of the degree of ionisation (middle panel). While linear fits are shown in the various panels, the purpose of this figure is mainly qualitative, and it highlights that a negative trend with galactocentric radius corresponds to a positive trend with metallicity (derived from the N2~ratio) and vice versa. A more quantitative study of the \HII\ regions based on photoionisation modeling will be discussed in a forthcoming publication, but we detail the fit parameters in Table~\ref{tab:fits}. The forthcoming quantitative \HII\ region study will also contain a detailed comparison between abundances derived via the strong line method from the MUSE data and corresponding values obtained via the direct temperature method from \citeauthor{bresolin09}. These types of comparisons are crucial for refining strong line calibrations \citep[e.g.,][]{curti17}, in particular given the rise of spatially resolved observations of orders of magnitude more regions in nearby galaxies. 

\begin{table}
\begin{center}
\caption{Slopes and intercepts obtained for the linear fits shown in Fig.~\ref{fig:gradients}, where the general form $f(x) = a \, x + b$ applies.}
\begin{tabular}{lcc}
\hline 
\hline
Parameter & $a$ & $b$ \\
\hline
$\log(\mathrm{S}32)$ & $0.49\pm0.22$ & $-0.96\pm0.06$ \\
$\log([\OIII]/\hb)$ & $1.88\pm0.30$ & $-0.79\pm0.08$ \\
$\log(\eta_{\NII})$ & $-0.61\pm0.27$ & $-0.56\pm0.08$ \\
\hline
\hline
\label{tab:fits}
\end{tabular}
\end{center}
\end{table}

\subsection{Feedback-related pressure terms}
\label{sec:press}

With the picture emerging from the previous section in which lower-metallicity \HII\ regions tend to have harder radiation fields, we now ask the question of whether this has measurable consequences on the impact of stellar feedback in these regions. To assess stellar feedback from the optical IFU data, we proceed as in \citetalias{mcleod20} in computing feedback-related pressure terms (see also \citealt{lopez14} and \citealt{olivier20}). Specifically, we focus here on the pressure of the ionised gas, $P_{\HII}$, and the direct radiation pressure, $P_\mathrm{dir}$, but do not quantify the effect of stellar winds. Compared to the pressure of the ionised gas, stellar winds have been shown to be increasingly less effective towards evolved \HII\ regions \citep[e.g.][]{lopez14, barnes20, mcleod20}, and they are expected to be weaker at lower metallicities \citep{kudritzki02}, such that the ionised gas pressure will therefore likely dominate the total feedback pressure, in particular at \HII\ region radii $\gtrsim$10 pc like the ones observed here . In \citetalias{mcleod20}, we demonstrate that stellar wind pressures of NGC~300 \HII\ regions can be derived from the resolved population of O-type stars in the regions, and we are currently working towards identifying individual massive stars from the MUSE data to then be able to quantify the relative importance of stellar winds and ionisation towards total \HII\ region pressures as well. Further, even though $P_\mathrm{dir}$ is only marginally contributing to the current expansion of the \HII\ regions observed here (as is shown later in this section), its impact during the embedded and compact stages of the regions' evolution contributed towards the present day region properties and is therefore certainly worth investigating.

The ionised gas pressure is computed from the electron density, $n_{\mathrm{e}}$, as
\begin{equation}
P_{\HII}=(n_{\mathrm{e}}+n_{\mathrm{H}}+n_{\mathrm{He}})kT_{\mathrm{e}}\approx2n_{\mathrm{e}}kT_{\mathrm{e}}~,
\label{pion}
\end{equation}
where we assume an \HII\ region temperature of $T_\mathrm{e} = 10^{4}$~K as well as singly-ionised helium. With \HII\ region temperatures in NGC~300 ranging between ${\sim}8000$~K and ${\sim}12000$~K \citep[][]{bresolin09}, this is a reasonable assumption. The direct radiation pressure is evaluated from the combined (bolometric) luminosity, $L_\mathrm{bol}$, of the massive stellar population in each region,
\begin{equation}
P_{\mathrm{dir}} = \frac{3L_\mathrm{bol}}{4\pi r^{2} c}~,
\label{pdir}
\end{equation}
where $r$ is the \HII\ region radius (which is derived from {\sc astrodendro} as the geometric mean of the major and minor axes of the projection onto the position--position plane, i.e.\ of an ellipse that describes the identified region, and for which we assume a measurement uncertainty of 20~per~cent). To compute the bolometric luminosity, we assume that the IMF and stellar age distribution are fully sampled in the \HII\ regions and convert the extinction-corrected H$\alpha$ luminosity to the bolometric luminosity according to \citet{kennicutt12}, $L_\mathrm{bol} = 138 L(\ha)$. As discussed in \citet{lopez14}, there are caveats to the conversion from H$\alpha$ to bolometric luminosity for young star-forming regions and stellar populations. First, because we are likely tracing stellar populations that are younger than ${\sim}5$~Myr, for which the ratio of H$\alpha$ to bolometric luminosity is higher than the \citeauthor{kennicutt12} value, we are likely overestimating the bolometric luminosity, and this overestimation will be different for regions of different ages. Second, the assumption of fully sampling the IMF is no longer valid for star clusters with masses below ${\sim}10^{4}$~M$_{\odot}$, introducing stochastic variations of the H$\alpha$ to bolometric luminosity ratio, which for a randomly selected cluster would imply an underestimation of $L_\mathrm{bol}$ \citep[e.g.][]{krumholz15,haydon20}. Similar to \citet{lopez14}, the selection of the \HII\ region sample investigated here is not random but based on H$\alpha$ emission, mitigating the stochastic effect to a factor ${\sim}2$ in the level of uncertainty. 

\begin{figure*}
    \centering
    \includegraphics[scale=0.6]{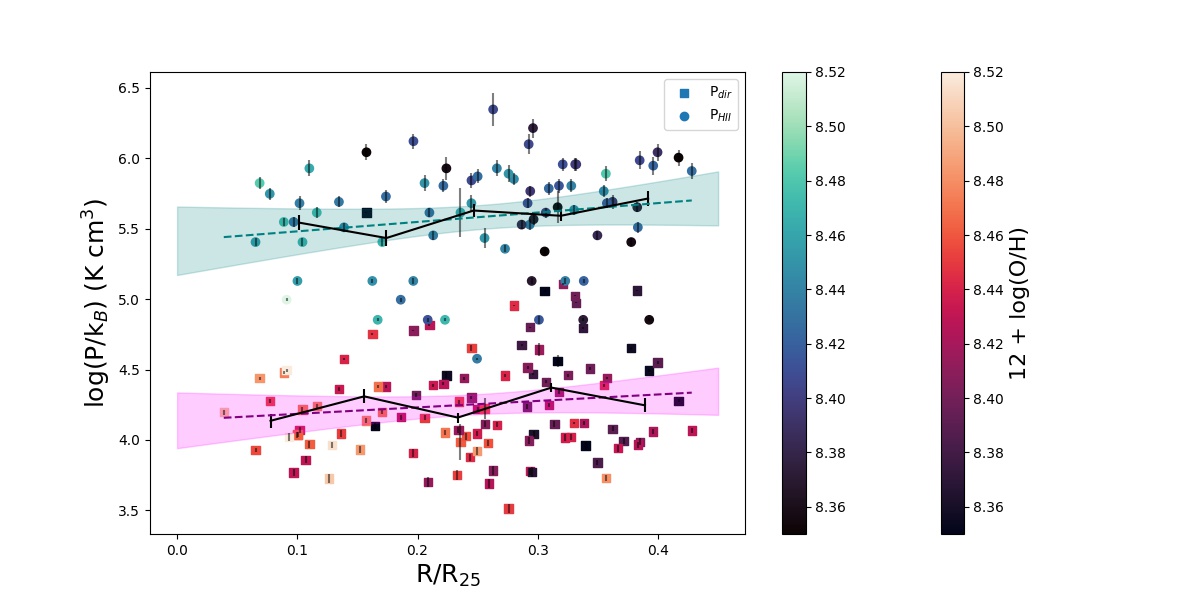}
    \caption{Ionised gas pressure, $P_{\HII}$, and direct radiation pressure, $P_\mathrm{dir}$, in \HII\ regions as a function of galactocentric radius and colour-coded by gas-phase abundance. Dashed lines are linear fits to the respective data points (and shaded areas are 95\% confidence regions): $\log(P_{\HII}/k_\mathrm{B})$ [K~cm$^{-3}$] $ = 5.38\,(\pm0.12) + 0.76\,(\pm0.45) \, R/R_{25}$ and $\log(P_\mathrm{dir}/k_\mathrm{B})$ [K~cm$^{-3}$] $ = 4.09\,(\pm0.09) +  0.60\,(\pm0.36) \, R/R_{25}$. Solid black lines show the trend when binning the pressure terms into 5 radial bins. See Section~\ref{sec:press}.}    
    \label{fig:pressures}
\end{figure*}

Fig.~\ref{fig:pressures} explores the impact of environment on the feedback-related pressure terms computed above. As already found by \citet{lopez11}, \citet{mcleod19}, and in \citetalias{mcleod20}, the predominantly more evolved \HII\ regions studied here are dominated by the pressure of the ionised gas, and we find $P_{\HII}/P_\mathrm{dir} > 1$ (the median across all \HII\ regions is ${\sim}56$).
While a clear correlation cannot be drawn and correlation coefficients reveal weak relations at best given the large scatter (with correlation coefficients of ${\sim}0.20$ and $0.18$ for $P_{\HII}$ and $P_\mathrm{dir}$, respectively), both $P_{\HII}$ and $P_\mathrm{dir}$ tend to increase with increasing galactocentric radius, along the negative metallicity gradient. To further illustrate the trend we divide the regions into five radial bins, the mean value of which are plotted against galactocentric radius in Fig.~\ref{fig:pressures}, and these having correlation coefficients of ${\sim}0.72$ and ${\sim}0.64$ for $P_{\HII}$ and $P_\mathrm{dir}$, respectively. Here, we use the distance from the centre of NGC~300 as an indicator of the varying ISM conditions, given that we have shown in Section~\ref{sec:abun} that along with the gas-phase abundance, the degree of ionisation and the hardness of the radiation field also show radial trends. We do not, however, explicitly plot the pressure terms against these, because similar emission lines and line ratios go into computing the different quantities, thus contaminating any resulting relations. For example, $P_\mathrm{dir}$ and metallicity are both dependent on the H$\alpha$ flux, while $P_{\HII}$ depends on the [\SII] lines, as do $\eta_{\NII}$ and S32. Independent measurements are required to attempt to quantify the metallicity dependence of $P_\mathrm{dir}$ and $P_{\HII}$, for example by obtaining deeper \HII\ region spectra and derive abundances from the direct method, by using different line ratios, or by combining emission line measurements with photoionisation modeling. 

\begin{figure}
    \centering
    \includegraphics[scale=0.46]{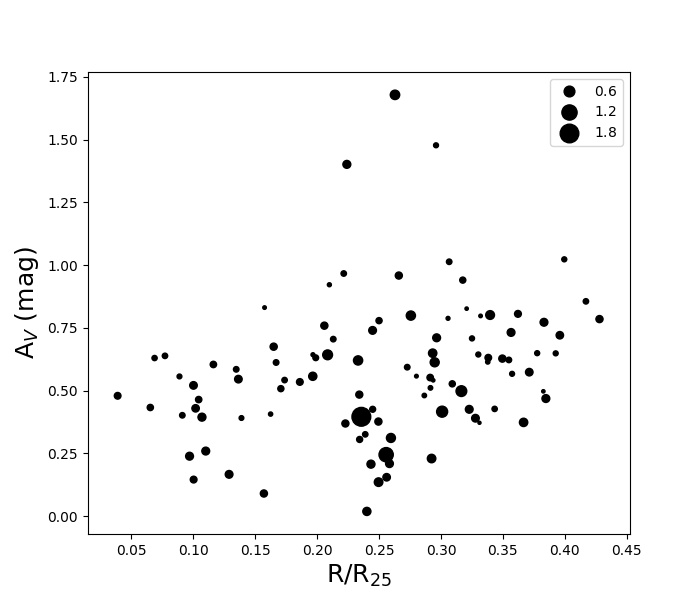}
    \caption{\HII\ region extinction, $A_{V}$, derived from the Balmer decrement, as a function of galactocentric radius. The sizes of the data points reflect the error on the $\ha/\hb$ ratio. \HII\ regions with unrealistic negative $A_{V}$ values do not appear in this figure.}
    \label{fig:av}
\end{figure}

An increase in $P_{\HII}$ with decreasing metallicity can be explained with the harder ionising radiation and increased photon fluxes. Further, while we have assumed a fixed temperature for all \HII\ regions, the negative metallicity gradient in NGC~300 leads to a positive temperature gradient due to line cooling being less efficient at lower metallicities \citep[][]{bresolin09}. If we were to take this into account, the increase of $P_{\HII}$ towards larger galactocentric distances would be further enhanced. The increase of $P_\mathrm{dir}$ is not as easily understood. The well established relation between the gas-to-dust ratio (G/D) and metallicity shows that the dust content tends to decrease with decreasing metallicity \citep[e.g.][and references therein]{remy14}, and we would therefore expect a decrease in radiation pressure at lower metallicities due to less dust to impart momentum to. This is indeed the case when comparing $P_\mathrm{dir}$ values of \HII\ regions in the LMC and SMC studied by \citet{lopez14}: the SMC \HII\ regions (where the metallicity and dust content are lower than in the LMC, e.g.\ \citealt{Roman_Duval_2014}) have systematically lower radiation pressure values than the LMC regions. 
We assess the amount of dust towards the \HII\ regions in NGC~300 from the reddening correction, as {\sc pyneb} returns both the colour excess, $E(B-V)$, and the extinction, $A_{V}$, based on the measured $\ha/\hb$ ratios. Fig.~\ref{fig:av} reveals a positive \HII\ region extinction gradient in the part of NGC~300 covered by the MUSE mosaic, which is in agreement with \citet{casasola17} who found that the dust mass surface density in NGC~300 is slightly increasing up to about ${\sim}0.5 R_{25}$ (their fig.~A.2). At the larger galactocentric radii not covered by MUSE the dust mass surface density decreases \citep{casasola17}, and additional MUSE data probing regions beyond $0.45 R_{25}$ are required to provide additional insight on the dependence of $P_\mathrm{dir}$ on the dust content. We note that this does not go against the G/D--metallicity relation, given that the metallicity range probed by the \HII\ regions in the inner ${\sim}2.5$~kpc of NGC~300 is rather small, i.e.\ ${\sim}0.2$~dex, range in which the scatter around mean G/D values can be large \citep{galametz11, remy14} such that a correlation within the small metallicity range cannot be established. We suggest that the $P_\mathrm{dir} {-} R$ relation shown in Fig.~\ref{fig:pressures} is driven by the dust content, together with the fact that at lower metallicity result in higher photon fluxes. 

\begin{figure*}
    \centering
    \includegraphics[scale=0.6]{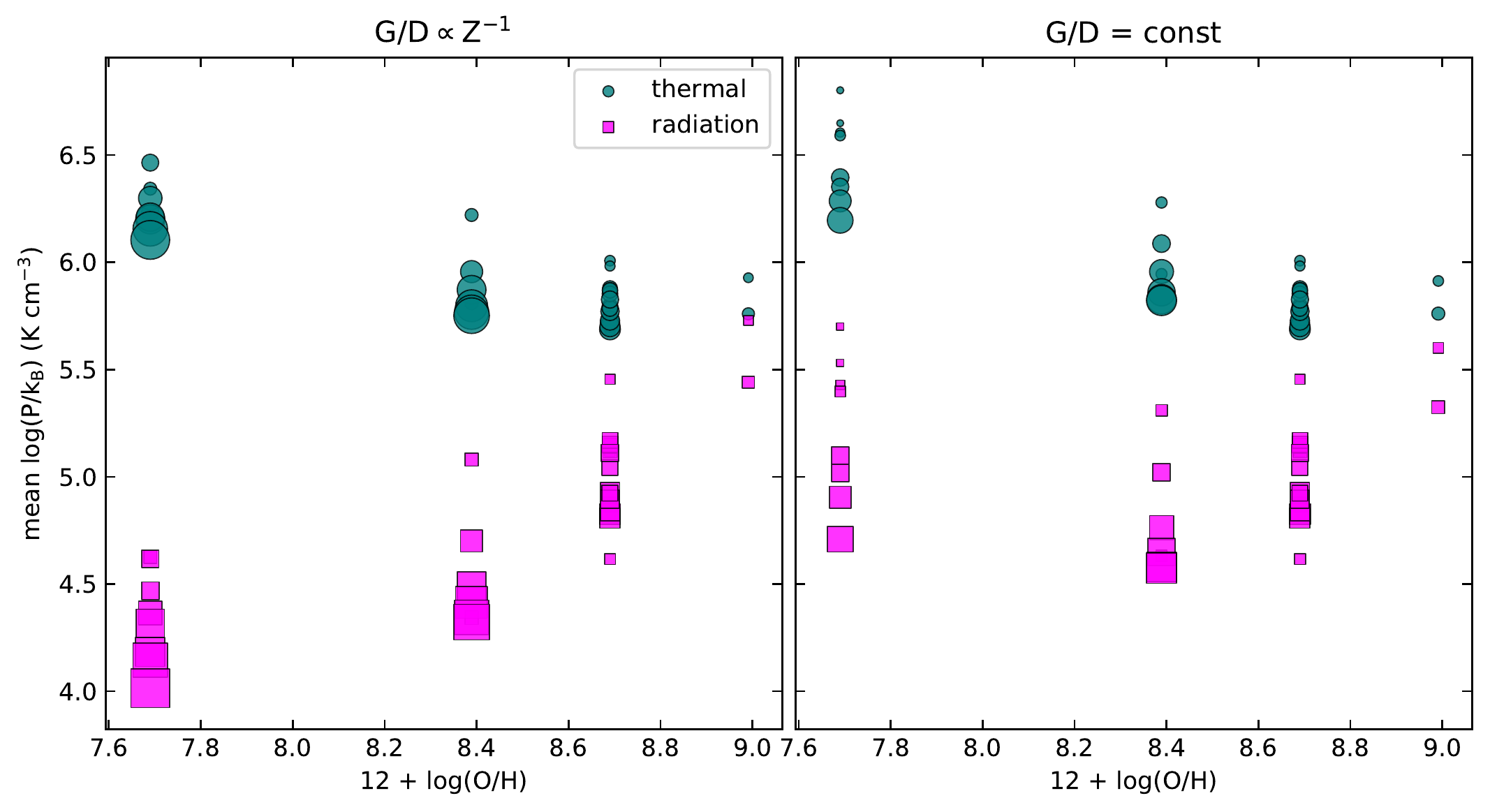}
    \caption{Mean thermal pressure and radiation pressure calculated from the radiation-hydrodynamical models of \citet{ali21}. \HII\ regions are extracted at different snapshots in time for each of the four different implemented metallicities. The left panel shows regions which evolved with gas-to-dust ratio inversely proportional to metallicity. The right panel shows post-processed snapshots using a single gas-to-dust ratio, thus mimicking a dust content trend similar to the one found in the considered portion of NGC~300. The marker size is proportional to the effective radius of the \HII\ region (defined as the radius of a sphere which would be equal to the same volume). }    
    \label{fig:models}
\end{figure*}

We further explore the role of metallicity and dust content in regulating pressure terms with the radiation-hydrodynamical models of \citet{ali21}. These models used the Monte Carlo radiative transfer code \textsc{torus} \citep{harries19} to explore stellar feedback in clusters at four different metallicities ($2, 1, 0.5, 0.1~\mathrm{Z}_{\odot}$). This method calculated photoionisation equilibrium and radiation pressure (direct plus dust-processed pressure) step-by-step with hydrodynamics in 3D. Fig.~\ref{fig:models}  shows the mean thermal pressure and radiation pressure in \HII\ regions extracted at different evolutionary times. The left panel shows results from the original models of \citet{ali21} in which the gas-to-dust ratio is proportional to metallicity ($\mathrm{G/D} \propto 100 / (Z/\mathrm{Z}_\odot)$), therefore serving as the control run where the dust content and the metallicity both decrease. The right panel shows post-processed snapshots where the radiative transfer has been recalculated with a fixed $\mathrm{G/D} = 100$ for all metallicities (i.e.\ boosting the dust content at sub-solar metallicity and lowering it at super-solar metallicity), therefore emulating environments similar to those observed in the covered portion of NGC~300 in terms of dust mass surface density. Together with the observations, these preliminary results indicate that the radial variation of the dust content is indeed likely playing the dominant role in regulating the $P_\mathrm{dir} {-} R$ relation: an increased amount of dust results in greater extinction of UV photons, making the \HII\ regions smaller and the radiation pressure larger. With simulations in general adopting the canonical G/D--metallicity relation, we therefore note that across spatially resolved scales within observed systems the dust content can in fact remain constant or even increase despite the presence of a negative metallicity gradient, which therefore can impact the relative importance of radiation pressure. This is likely even more important in young, compact \HII\ regions, and should be taken into account when simulations are tailored to match specific observed systems.

We note that the radial pressure trends observed here are however a combination of several intrinsic and environmental factors, e.g.\ gas/dust content, metallicity, star formation rate, evolutionary stage of the \HII\ regions, age of the stellar population, the relative importance of which needs to be assessed. For example, the star formation rate (a relative increase of which over a given period of time would result in overall increased stellar feedback) is observed to be approximately constant in the portion of the galaxy covered by our observations (out to about $0.45 R_{25}$; \citealt{gogarten10, williams13, casasola17}), and we therefore do not consider it to be a dominant source driving the increase of stellar feedback here. In terms of evolutionary stage, one would expect the radiation pressure to be enhanced in younger, more compact regions when compared to more evolved regions  \citep{olivier20}. With a radius cutoff of about $7$~pc as described we are not able to probe compact and ultra-compact \HII\ regions, and an evolutionary dependence as traced by the ages of the stellar populations within the regions will be explored in a forthcoming paper. Further, the wealth of existing and upcoming IFU data from nearby galaxy surveys like SIGNALS or PHANGS which contain targets with substantial ancillary, co-spatial multi-wave\-length coverage is ideal to study environmental dependencies of stellar feedback. Current substantial effort towards this will certainly lead to further insight in this area in the near future.

\section{Supernova remnants in the context of early stellar feedback}
\label{sec:preshock}

Supernova feedback has long been considered to be the main mechanism responsible for driving turbulence and regulating star formation rates and efficiencies on galaxy-wide scales. However, in recent years simulations have started to show that early, pre-SN feedback (radiation pressure, ionisation, stellar winds) need to be accounted for because the energy deposited by SNe is not sufficient to disrupt and disperse dense molecular clouds (see Section~\ref{sec:intro}). For NGC~300, we applied the statistical methodology of \citet{kruijssen14} and \citet{kruijssen18} to characterise the GMC lifecycle and found that the GMCs in this galaxy are dispersed on short time-scales of $1.5\pm0.2$~Myr, requiring pre-SN feedback \citep{kruijssen19}. This result has been generalised in recent observational studies by \citet{chevance20b,chevance20}, who have quantified feedback time-scales across nine nearby disc galaxies, finding that GMCs are dispersed within $1{-}5$~Myr after the emergence of massive stars from their dust-enshrouded birth places. This analysis has been extended further by \citet{kim21} to also encompass the dust embedded stages of star formation than those probed by \citeauthor{chevance20}, and both studies conclude that early (pre-SN) stellar feedback (in the form of stellar winds and photoionisation in particular) is a major component driving the GMC disruption. These results are consistent with optical/near-ultraviolet studies of the young cluster population, which find that clusters become unassociated with their natal clouds after just a few Myr \citep[e.g.][]{hollyhead15,grasha18,hannon19}. Both \citeauthor{chevance20} and \citeauthor{kim21} do not find significant correlations between environmental properties and the time-scales over which feedback acts.

As described in Section~\ref{sec:gas}, the \HII\ regions in the inner portion of NCG~300 reveal a trend of increasing pre-SN feedback (traced by the direct radiation pressure and the pressure of the ionised gas) with increasing galactocentric radius, i.e.\ along NGC~300's negative metallicity and positive extinction gradient. The massive stellar populations residing in the \HII\ regions are not only the drivers of early stellar feedback, but are also the progenitors of SN events. In other words, (core collapse) SNe typically occur in \HII\ regions within a few Myr of the onset of pre-SN feedback, which has already started to affect the surrounding ISM and alter the environment into which SNe expand into \citep[e.g.][]{haid18,lucas20,keller21b}.

Hence, we now ask the question of whether there are any systematic differences in terms of the environment of the seven detected SNRs, and if yes, what the likely driver of the environmental differences is. To probe the pre-SN environment we compute the pre-shock ISM density, $n_\mathrm{ISM}$ (the density of the ISM in which SNe went off into, which differs from the density as one would measure from, e.g.\ the ratio of the [\SII] lines, as this would deliver the electron density of the shocked matter), proceeding as in \citet{mcleod19}: We use the relation between the flux of the H$\beta$ line from a surface element of a radiative shock with velocity, $v_\mathrm{s}$, and ISM density, $n_\mathrm{ISM}$, as given in \citet{dopita96},
\begin{equation}\label{eq:dopita}
    \frac{f_{\hb}}{\mathrm{erg~s^{-1}~cm^{-2}}} = 7.44\times10^{-6} \, \Bigg(\frac{v_\mathrm{s}}{100~\mathrm{km}~\mathrm{s}^{-1}}\Bigg)^{2.41} \Bigg(\frac{n_\mathrm{ISM}}{\mathrm{cm^{-3}}}\Bigg)~.
\end{equation}
A bubble (e.g.\ a SNR) of radius~$r$ at a distance~$D$ from the observer spans a solid angle of $\Omega = \pi (r/D)^{2}$ on the sky and is observed with a flux $F_{\hb} = L/(4 \pi D^{2}) = 4 \pi r^{2} f_{\hb} / (4 \pi D^{2})$ and an intensity $I_{\hb} = F_{\hb}/\Omega = f_{\hb}/\pi$. Hence, Eq.~\ref{eq:dopita} becomes
\begin{equation}
    \frac{I_{\hb}}{\mathrm{erg~s^{-1}~cm^{-2}~sr^{-1}}} = 2.36\times10^{-6} \, \Bigg(\frac{v_\mathrm{s}}{100~\mathrm{km}~\mathrm{s}^{-1}}\Bigg)^{2.41} \Bigg(\frac{n_\mathrm{ISM}}{\mathrm{cm^{-3}}}\Bigg)~.
\end{equation}
With the shock velocities measured from the [\SII] FWHM of the SNRs ($v_\mathrm{s} \approx \mathrm{FWHM}$; \citealt{heng10}), we derive pre-shock ISM densities on the order of a few particles per cm$^{-3}$ (see Table~\ref{tab:snr_params}), values that are consistent with shocks in a pre-ionised medium \citep{dopita17}.

\begin{figure*}
    \centering
    \begin{subfigure}[b]{0.45\textwidth}
    \centering
         \includegraphics[scale=0.47]{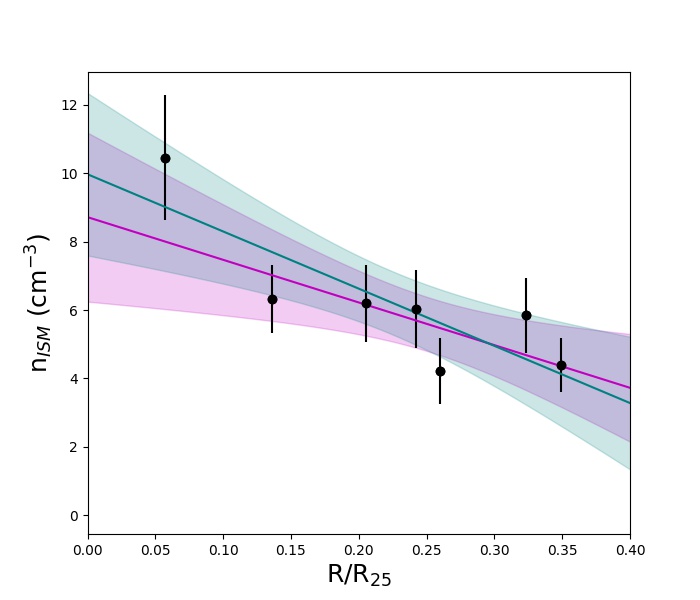}
         \caption{}
         \label{fig:rad_den}
    \end{subfigure}
    \hfill
     \begin{subfigure}[b]{0.45\textwidth}
         \centering
         \includegraphics[scale=0.47]{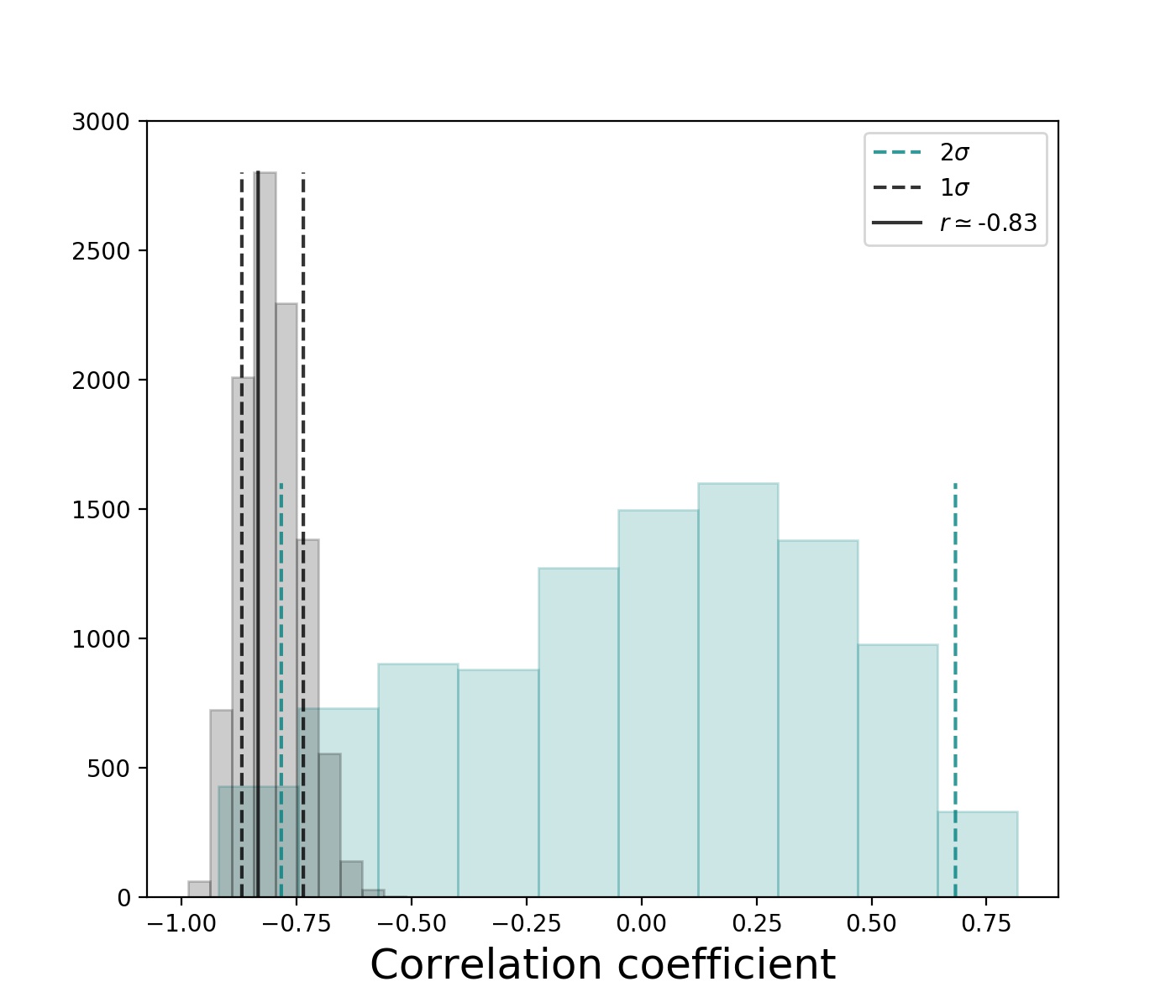}
         \caption{}
         \label{fig:pear_rad}
     \end{subfigure}
     \hfill
         \begin{subfigure}[b]{0.45\textwidth}
    \centering
         \includegraphics[scale=0.47]{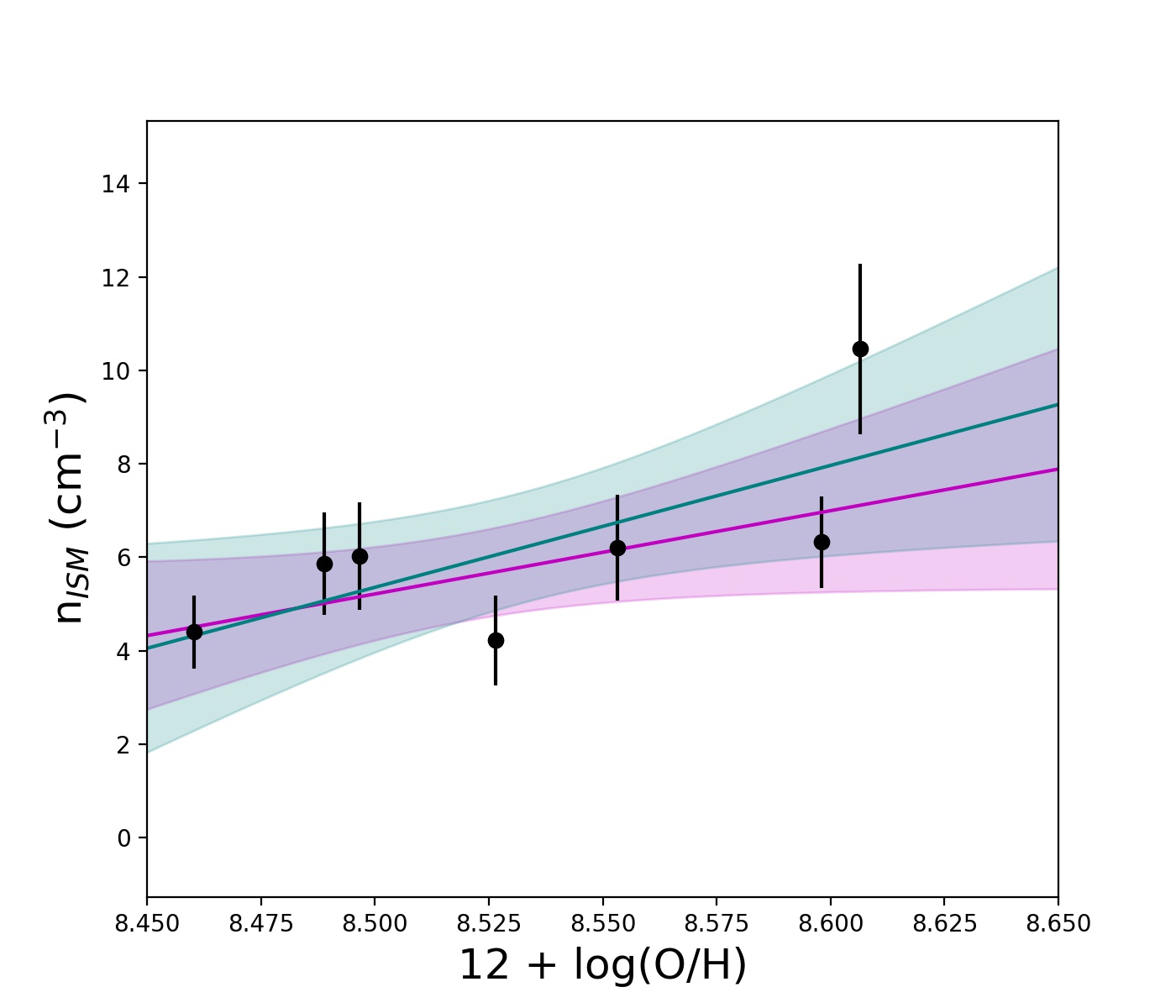}
         \caption{}
         \label{fig:ox_den}
    \end{subfigure}
    \hfill
     \begin{subfigure}[b]{0.45\textwidth}
         \centering
         \includegraphics[scale=0.47]{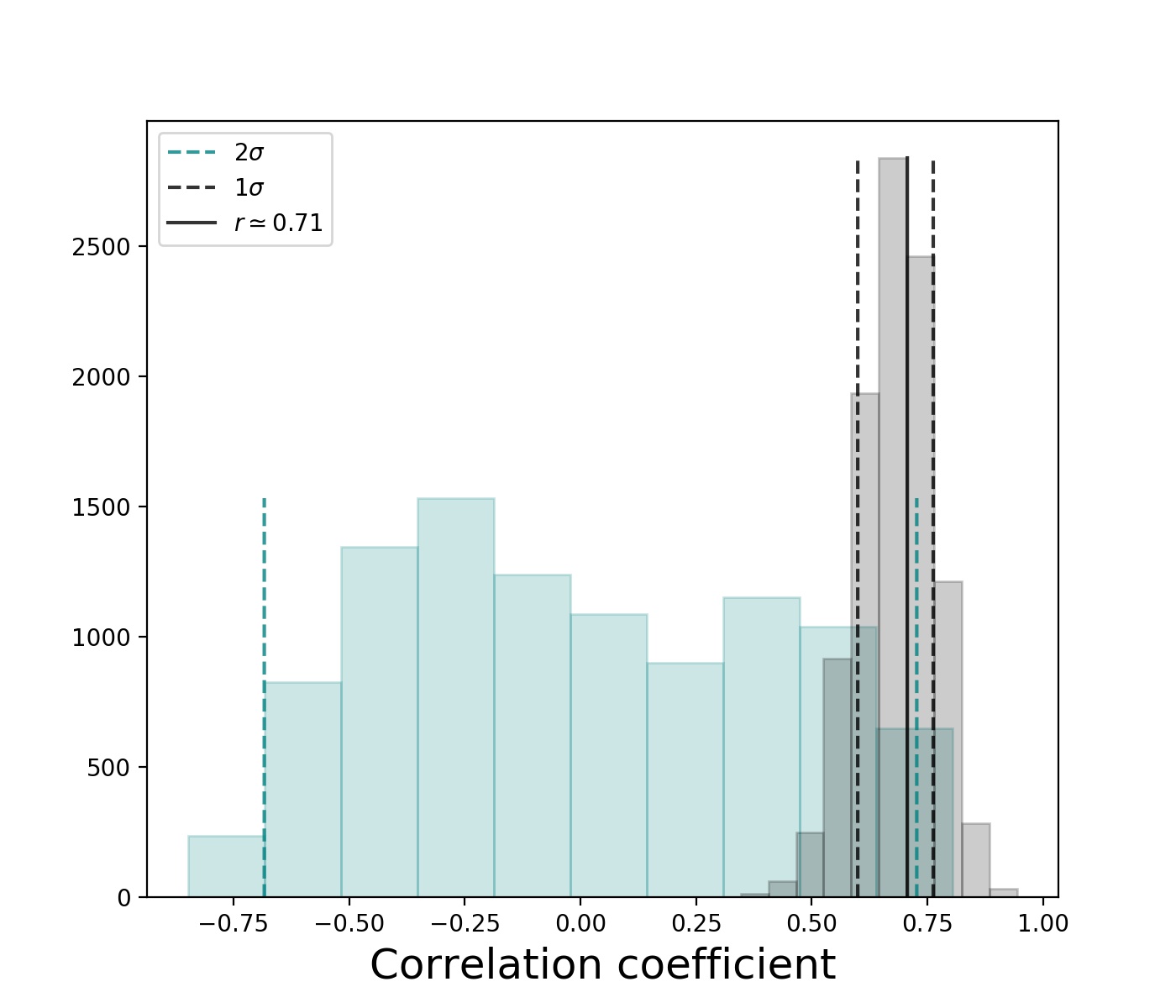}
         \caption{}
         \label{fig:pear_ox}
     \end{subfigure}
     \caption{Panel~\subref{fig:rad_den}: pre-shock ISM density as a function of galactocentric radius~$R$. The magenta and teal lines are linear fits with and without uncertainties, respectively, together with 95~per~cent confidence regions. Panel~\subref{fig:pear_rad}: correlation coefficient analysis for the $n_\mathrm{ISM}{-}R$ relation (see Section~\ref{sec:preshock}, $r \simeq -0.83$, solid black line); the black histogram is the result of varying $n_\mathrm{ISM}$ within the respective error bars, and the teal histogram results from randomising~$R$, dashed vertical lines correspond to $1\sigma$ and $2\sigma$ ranges, respectively. Panel~\subref{fig:ox_den}: pre-shock ISM density as a function of gas-phase abundance, the colours are the same as in panel~\subref{fig:rad_den}. Panel~\subref{fig:pear_ox}: same as panel~\subref{fig:pear_rad}, but for the $n_\mathrm{ISM}$--abundance relation ($r \simeq 0.71$). \label{fig:preshock}}
\end{figure*}

With an estimate of the pre-shock ISM density for each SNR, we now explore whether there is any trend with galactocentric radius and/or metallicity, thus testing the scenario in which the enhanced early stellar feedback at these galactocentric radii (i.e.\ metallicities) could have created less dense environments in which subsequently the supernova events occurred. Fig.~\ref{fig:rad_den} shows the inferred pre-shock ISM density, $n_\mathrm{ISM}$, as a function of galactocentric radius, together with two linear fits (with and without taking into account the uncertainties on $n_\mathrm{ISM}$ in magenta and orange, respectively). While there appears to be a relation in the expected direction, the limited number of SNRs in the MUSE mosaic is likely hindering a more robust interpretation. To further explore the significance (or lack thereof) of the relation between $n_\mathrm{ISM}$ and~$R$, we compute the Pearson correlation coefficient~$r$ which, for the $n_\mathrm{ISM}{-}R$ values shown in Fig.~\ref{fig:preshock}\subref{fig:rad_den} is about $-0.83$, indicating a relatively strong negative relation. We assess the uncertainty of $r$ by bootstrapping while: (i) randomising $R$ and (ii) varying the $n_\mathrm{ISM}$ values within their respective uncertainties.

With (i) we test whether reshuffling the $x$-axis yields the original relation between $n_\mathrm{ISM}$ and~$R$, hence testing the null-hypothesis of there being no relation between $n_\mathrm{ISM}$ and~$R$. By bootstrapping this with $10^{4}$ iterations, we thus obtain a statistical significance of nearly $3\sigma$ for the relation. Or, in other words, by randomising $R$ we do not recover the relation $99.64$~per~cent of the times.
With (ii) we quantify the relation given the uncertainties on $n_\mathrm{ISM}$, thus computing the uncertainty of~$r$. The correlation coefficient analysis for the $n_\mathrm{ISM}{-}R$ relation is shown in Fig.~\ref{fig:preshock}\subref{fig:pear_rad}, where the black histogram corresponds to randomising~$R$, while the teal histogram corresponds to varying $n_\mathrm{ISM}$: within $1\sigma$ of~$r$, the correlation is $2\sigma$ away from the null-hypothesis representing no relationship. We further analyse the environmental dependency of the pre-shock ISM density in Fig.~\ref{fig:preshock}\subref{fig:ox_den} and Fig.~\ref{fig:preshock}\subref{fig:pear_ox}, which show that a corresponding positive correlation is found with metallicity, as would be expected if increased pre-SN feedback at lower metallicities created lower density environments. While the correlation between the pre-shock ISM density and the gas-phase metallicity, $r \simeq 0.71$, is not as strong as the one with galactocentric radius above ($r \simeq -0.83$), we still recover a $2\sigma$ significance within $1\sigma$ of~$r$.

These findings are in excellent agreement with the results of \citet{lucas20}, who simulate supernovae in star-forming molecular clouds with and without early stellar feedback (in the form of ionisation and stellar winds). In their study, \citeauthor{lucas20} find that early stellar feedback creates pre-SN cavities with systematically lower densities. Further, compared to the control run, they find that including pre-SN feedback leads to simulated clouds with enhanced low column density channels via which the SN ejecta (and shock-heated gas) can escape more easily, thus affecting the lower density ISM at greater distances from the natal cloud. This means that because of early stellar feedback, SNe can deposit more energy on galactic scales. Conversely, \citet{smith21} find that including early stellar feedback in their simulations disrupts molecular clouds and leads to less SN clustering, reducing outflow rates in a substantial manner. \citet{keller21b} apply an empirically-motivated, early feedback model based on the observations of \citet{kruijssen19} and \citet{chevance20b,chevance20} in isolated disc galaxy simulations, and also find that the inclusion of early feedback reduces SN clustering, but next to a moderate decrease of the time-averaged outflow, its burstiness drops precipitously, leading to a considerably `smoother' galaxy-scale baryon cycle. While our findings do not allow us to quantify the impact of the SNe themselves, these numerical studies clearly demonstrate the importance of the measurement in Fig.~\ref{fig:preshock}.

The environmental dependencies found here are consistent with the results of \citet{chevance20} and \citet{kim21}, who observed no statistically significant radial trends for the derived feedback time-scales within the uncertainties. Indeed, the increase of feedback pressure reported here does not directly imply accelerated feedback time-scales, as the time-scales depend on other environmental properties that set how rapidly the imparted feedback (which is what we measure) disperses the natal GMCs. In fact, our findings support the weak relation between metallicity and feedback time-scales found by \citeauthor{chevance20b} where lower metallicity galaxies have shorter feedback time-scales. Furthermore, this underpins the results of \citet{kruijssen19}, who showed that the GMC lifetimes and feedback time-scales both decrease with increasing galactocentric radius in NGC~300 and that these are regulated by early stellar feedback. An accelerated feedback timescale towards larger galactocentric radii would imply that one would observe less evolved regions at larger radii, which could in turn contribute to the radial increase of $P_\mathrm{dir}$ as it is enhanced in younger regions.

We caution that while we show direct evidence for an environmental dependency of the pre-shock ISM density in the form of galactocentric radius and metallicity, a causal connection between enhanced $P_\mathrm{dir}$ and $P_{\HII}$ and lower pre-shock ISM densities, while tempting, cannot be directly made from the observations, so that this connection is purely inferred. 
We also stress that while the correlation coefficient analysis indicates that the $n_\mathrm{ISM}$--environment relation is relatively robust, additional MUSE coverage of more SNRs in NGC~300 at larger radii and lower metallicities would be needed to further strengthen the result. Two additional points need to be made. First, while the SNR identification is robust, the sample is potentially missing old, evolved SNRs with low surface brightness and unresolved shock velocities. Second, with the focus here being on SNRs, this analysis does not allow the investigation of SNe that potentially already occurred within the \HII\ regions but are not identified here due to the dominating \HII\ region spectra. More clarity towards this will be obtained by analysing the age of the stellar populations within the regions.

\begin{table*}
\begin{center}
\caption{Properties of SNRs (see text Section~\ref{sec:preshock}). Columns correspond to the dendrogram id (1), the galactocentric radius (2), the H$\beta$ intensity (3), the inferred shock velocity (4), the pre-shock ISM density (5) and the SNR gas-phase abundance (6).}
\begin{tabular}{cccccc}
\hline 
\hline
Dendrogram id & $R$ & $I(\hb)$ & $v_{2}$ & $n_\mathrm{ISM}$ & 12+log(O/H) \\
\hline
 & (R$_{25}$) & (10$^{-5}$ erg s$^{-1}$ cm$^{-2}$ sr$^{-1}$) & (100 km s$^{-1}$) & (cm$^{-3}$) & \\
\hline
106  &  0.32  &  6.61 $\pm$ 0.20  &  1.91 $\pm$ 0.15  &  5.85 $\pm$ 1.10  &  8.49 $\pm$ 0.01\\
339  &  0.14  &  7.19 $\pm$ 0.17  &  1.92 $\pm$ 0.12  &  6.32 $\pm$ 0.98  &  8.60 $\pm$ 0.01\\
529  &  0.06  &  7.06 $\pm$ 0.19  &  1.55 $\pm$ 0.11  &  10.46 $\pm$ 1.82  &  8.61 $\pm$ 0.01\\
543  &  0.21  &  4.97 $\pm$ 0.14  &  1.66 $\pm$ 0.12  &  6.20 $\pm$ 1.13  &  8.55 $\pm$ 0.01\\
662  &  0.35  &  3.03 $\pm$ 0.08  &  1.56 $\pm$ 0.11  &  4.40 $\pm$ 0.78  &  8.46 $\pm$ 0.01\\
721  &  0.24  &  4.88 $\pm$ 0.12  &  1.67 $\pm$ 0.13  &  6.03 $\pm$ 1.15  &  8.50 $\pm$ 0.01\\
788  &  0.26  &  4.00 $\pm$ 0.10  &  1.78 $\pm$ 0.17  &  4.22 $\pm$ 0.96  &  8.53 $\pm$ 0.01\\
\hline
\hline
\label{tab:snr_params}
\end{tabular}
\end{center}
\end{table*}

\section{Summary and conclusions}
\label{sec:conclusion} 

In this paper, we have used optical IFU data from the VLT/MUSE instrument to identify and classify emission line regions, and study the environmental properties of \HII\ regions and SNRs. The emission line region identification procedure exploits dendrograms, which are complemented with a spectral clustering algorithm which ensures that the spatially coherent structures are not over-fragmented. The main drawback is, like it is the case for most region identification methods, that the algorithm can sometimes group together into a single complex what are likely separate regions, particularly where several emission line regions are crowded together and spatially overlap. Given that we are mostly analysing radial dependencies, grouping together neighbouring regions only has a marginal impact on the analysis performed here. 

We separate regions into SNRs, PNe, and \HII\ regions based on emission line ratios and emission line widths, and show that this method is robust and can be readily used for nearby galaxy data sets of similar spectral and spatial coverage. We do however recommend that if the aim is to specifically identify SNRs and PNe, additional emission line maps to the H$\alpha$ one should be used (e.g.\ [\SII] for SNRs and [\OIII] for PNe) to ensure that objects with low H$\alpha$ surface brightness are also recovered. We stress that this method is purely empirical, and a comparison with other methods is planned for the future.

For the identified \HII\ regions we then derive several quantities associated with the ionised gas within them. These are the (oxygen) abundance, the degree of ionisation, and the hardness of the radiation field, and we recover the known negative metallicity gradient of the galaxy. We show that, within the portion of the galaxy covered by the MUSE mosaic (i.e.\ out to ${\sim} 0.45 R_{25}$), the degree of ionisation increases and the radiation fields become harder with increasing galactocentric radius and decreasing metallicity. This is consistent with the stellar populations residing in the \HII\ regions having higher photon fluxes and higher effective temperatures at lower metallicities, as inferred from higher degrees of ionisation and harder radiation fields. We then compute feedback-related pressure terms for ionisation and radiative feedback and discuss their dependence but also impact on their environment. With the strength of early stellar feedback seemingly increasing as a function of distance from the galactic centre and decreasing metallicity, we then analyse the environment into which SNe in NGC~300 occurred.

The main results from this paper are summarised as follows:
\begin{itemize}
\item We present a simple empirical method to separate \HII\ regions, PNe, and SNRs in nearby galaxy optical IFU observations. This method is based on three parameters only, namely the $[\OIII]/\hb$ ratio, the $[\SII]/[\OIII]$ ratio, and the width of the [\SII] line. While this method requires spectroscopic observations, it otherwise only relies on ratios of strong lines and circumvents assumptions on the IMF and uncertainties stemming from distance measurements. 
\item Trends with galactocentric radius and metallicity are found for the degree of ionisation (as traced by $[\OIII]/\hb$ and $[\SIII]/[\SII]$) and the hardness of the \HII\ region ionisation fields, such that at lower metallicities (i.e.\ larger galactocentric radii) we find harder radiation fields and increased degrees of ionisation indicating stars with higher effective temperatures and higher ionising photon fluxes. Both photoionisation modeling and auroral line measurements are needed to further disentangle intrinsic dependencies of the used strong line ratios to characterise the ionised gas.
\item Linked to the previous point, we observe weakly increased pre-SN feedback (traced by the direct radiation pressure and the pressure of ionised gas) in \HII\ regions at larger galactocentric radii and lower metallicities.
\item We suggest that the increased radiation pressure in \HII\ regions at larger galactocentric radii and lower metallicities is likely due to an increase in the dust content towards the outer regions probed by the data. Additional MUSE observations of the outer parts of NGC~300 (beyond the presently available coverage which extends to ${\sim} 0.45 R_{25}$) are needed to further quantify the dependence of $P_\mathrm{dir}$ on the dust content.
\item Preliminary results from dedicated simulations of star-forming molecular clouds support our conclusion that dust regulates the relative importance of radiation pressure. We suggest that, where available, the known dust mass surface density should be taken into account if simulations are tailored to reproduce specific galactic systems.
\item Given the radial and metallicity trends of pre-SN feedback we further study the density of the environment the detected SNR expanded into, and find that SNe at larger galactocentric radii (and at lower metallicities) expanded into lower-density environments. We tentatively suggest that this could be a consequence of the increased pre-SN feedback at low metallicities, which has contributed to creating lower-density environments, in excellent agreement with simulations. 
\end{itemize}

In conclusion, we reiterate that ongoing and upcoming IFU surveys of nearby galaxies enable \HII\ region and SNR studies analogous to what is presented in this paper, while providing a much wider range of different environmental properties to explore. These will then provide an observational quantification of the environmental impact and dependence of stellar feedback, which can be used to improve feedback prescriptions in simulations. 

\section*{Acknowledgements}

This research made use of astrodendro, a Python package to compute dendrograms of Astronomical data (http://www.dendrograms.org/). This research also made use of {\sc SCIMES}, a Python package to find relevant structures into dendrograms of molecular gas emission using the spectral clustering approach. AFM was partially supported by NASA through the NASA Hubble Fellowship grant No.~HST-HF2-51442.001-A awarded by the Space Telescope Science Institute, which is operated by the Association of Universities for Research in Astronomy, Incorporated, under NASA contract NAS5-26555. The \textsc{torus} models were performed using the DiRAC Data Intensive service at Leicester, operated by the University of Leicester IT Services, which forms part of the STFC DiRAC HPC Facility (www.dirac.ac.uk). The equipment was funded by BEIS capital funding via STFC capital grants ST/K000373/1 and ST/R002363/1 and STFC DiRAC Operations grant ST/R001014/1. DiRAC is part of the National e-Infrastructure. MC and JMDK gratefully acknowledge funding from the Deutsche Forschungsgemeinschaft (DFG, German Research Foundation) through an Emmy Noether Research Group (grant number KR4801/1-1) and the DFG Sachbeihilfe (grant number KR4801/2-1), as well as from the European Research Council (ERC) under the European Union's Horizon 2020 research and innovation programme via the ERC Starting Grant MUSTANG (grant agreement number 714907). AAA acknowledges funding from the European Research Council for the Horizon 2020 ERC consolidator grant project ICYBOB (grant number 818940).  AFM thanks S.~Scaringi, K.~Long, and W.~Blair for the useful discussions.

\section*{Data availability}

The data underlying this article will be shared on reasonable request to the corresponding author.




\bibliographystyle{mnras}
\bibliography{refs} 




\appendix

\section{Abundances from strong line ratios}
\label{app:abun}

In Section~\ref{sec:abun}, we derive strong line oxygen abundances for the detected \HII\ regions. For this we use the N2~ratio and the \citet{marino13} calibration. However, the N2~ratio itself has a slight dependence on the ionisation parameter, and here we explore the difference between using the empirical \citeauthor{marino13} calibration and using the theoretical calibration by \citet{kewley19}, which corrects for the ionisation parameter dependence. In addition, we also compare these two to the \citet{marino13} O3N2 and the \citet[][]{pilyugin16} calibrations.

\begin{figure}
    \centering
    \includegraphics[scale=0.47]{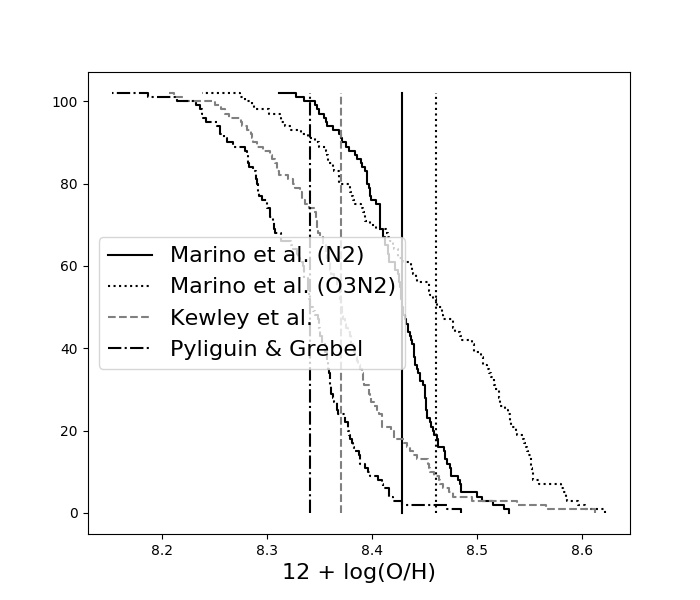}
    \caption{Cumulative distributions of the gas-phase abundance obtained from the N2 and O3N2 ratios as per the empirical \citet{marino13} calibration (solid and dotted black, respectively), the theoretical \citet{kewley19} calibration (grey), and the \citet[][]{pilyugin16} calibration (black dot-dashed). Mean values ($8.43$, $8.46$, $8.37$, and $8.34$, respectively) are indicated with vertical lines.}
    \label{fig:ox_compare}
\end{figure}

Fig.~\ref{fig:ox_compare} shows the cumulative distributions of the oxygen abundance derived from the two N2 calibration, for which a KS test yields a $p$-value of $1.4\times10^{13}$. Compared to \citeauthor{marino13}, both the \citeauthor{kewley19} and the \citeauthor{pilyugin16} diagnostics lead to systematically lower abundances, as well as a larger spread in values, leading to a significantly worse agreement with the temperature-based abundance gradients of \citet{toribio16} and \citealt{bresolin09}.

\begin{figure*}
    \centering
    \includegraphics[scale=0.5]{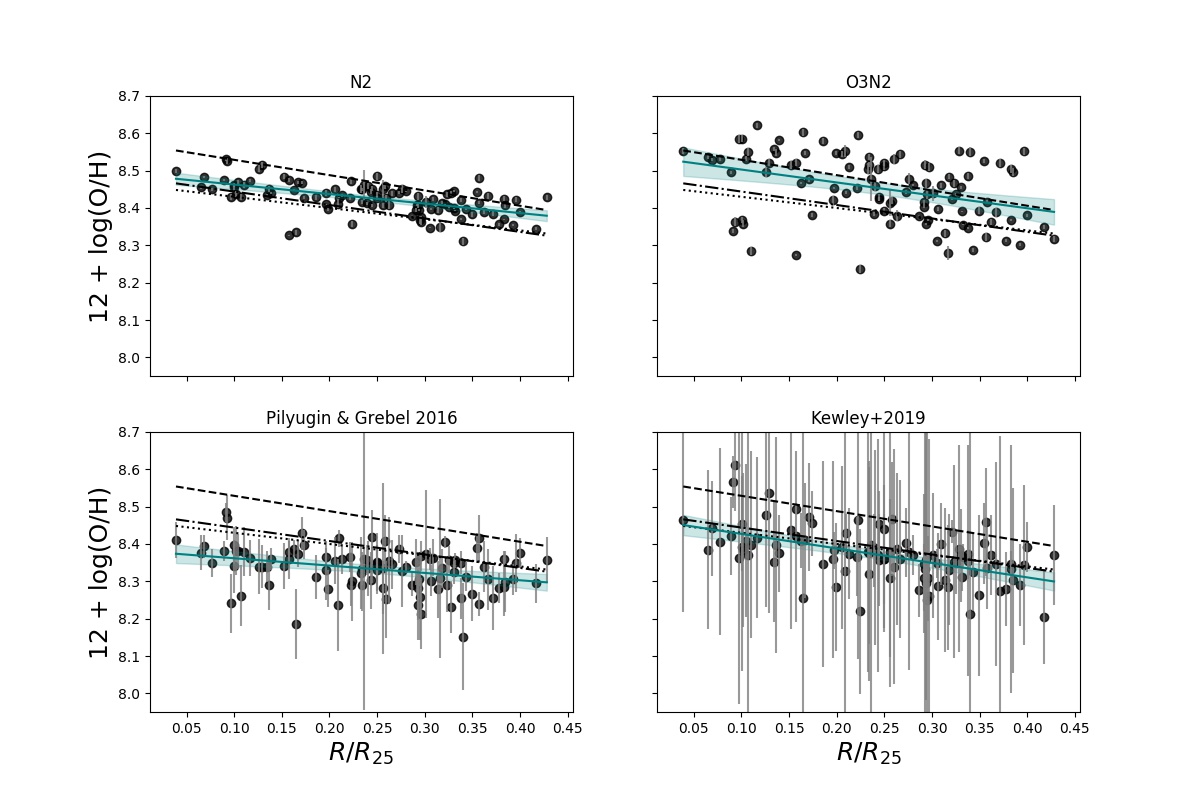}
    \caption{Metallicity gradients obtained from different diagnostics (see also Fig.~\ref{fig:metgrad}). In all panels the dotted, dashed, and dot-dashed lines are the gradients determined via the direct temperature-based method by \citet{toribio16}, \citet{bresolin09}, and \citet{stasinska13}, respectively.}
    \label{fig:ox_compare2}
\end{figure*}

The corresponding abundance gradients are shown in Fig.~\ref{fig:ox_compare2}, where the dotted, dashed, and dot-dashed lines are the gradients determined via the direct temperature-based method by \citet{toribio16}, \citet{bresolin09}, and \citet{stasinska13}, respectively (see Section \ref{sec:abun}). All of the diagnostics in the figure show the negative abundance gradient, albeit with different slopes and intercept. The N2 calibration shows the best agreement with the abundance gradients from the literature, as well as the least amount of scatter (see also Fig.~\ref{fig:ox_compare}). The larger errors in the bottom two panels reflect the larger uncertainties stemming from the fainter [SII] and [SIII] lines.

\section{Emission line region properties}
\label{app:tables}

In the following tables, we report the properties of the identified SNRs (Table~\ref{tab:snr}), PNe (Table~\ref{tab:pne}), and \HII\ regions (Table~\ref{tab:hii}).

\begin{table*}
\begin{center}
\caption{SNR table.}
\begin{tabular}{lcccccl}
\hline 
\hline
Millar12 id & Dendrogram id & Coordinates & log([\NII]/H$\alpha$) & log($[\SII]/\ha$) & FWHM$_{[\SII]}$ & Notes\\
\hline
 & (this work) & (J2000) & & & (\AA) & \\
\hline
S13 & 543 & 13.69415 -37.66244 & -0.41$\pm$0.02 & 0.07$\pm$0.02 & 3.72$\pm$0.34 &\\ 
S14 & 339 & 13.69633 -37.68562 & -0.32$\pm$0.02 & 0.11$\pm$0.02 & 4.30$\pm$0.33 &\\
S15 & 721 & 13.72290 -37.64731 & -0.53$\pm$0.01 & -0.05$\pm$0.01 & 3.73$\pm$0.36 &\\
S16 & 529 & 13.72795 -37.67664 & -0.30$\pm$0.02 & 0.11$\pm$0.02 & 3.46$\pm$0.31 &\\
S17 & 106 & 13.73577 -37.73257 & -0.55$\pm$0.02 & 0.04$\pm$0.20 & 4.29$\pm$0.40 &\\
S18 & 788 & 13.75687 -37.65529 & -0.47$\pm$0.01 & -0.13$\pm$0.01 & 3.99$\pm$0.46 &\\
S24 & 662 & 13.78971 -37.67494 & -0.61$\pm$0.02 & 0.02$\pm$0.01 & 3.49$\pm$0.31 &\\
\hline
S10 & - & 13.67029 -37.68020 & - & - & - & microquasar\\
S19 & - & 13.77254 -37.68918 & - & - & - & in \HII\ region contour\\
S22 & - & 13.78125 -37.67866 & - & - & - & below detection threshold\\
\hline
- & 538 & 13.72379 -37.67468 & -0.24$\pm$0.01 & 0.10$\pm$0.01 & 3.80$\pm$0.30 & SNR candidate\\
\hline
\hline
\label{tab:snr}
\end{tabular}
\end{center}
\end{table*}

\begin{table*}
\begin{center}
\caption{Parameters of PNe. Column 1 gives the id from \citet{stasinska13}, Column 2 the identifier from the dendrogram identification algorithm, Column 3 lists central coordinates, and Column 4 and 5 give line ratio values for each PN. Additional notes for some PNe are given in Columns 6.}
\begin{tabular}{lccccl}
\hline 
\hline
Stasi\'{n}ska13 id & Dendrogram id & Coordinates & log([\NII]/H$\alpha$) & log([\OIII]/H$\beta$) & Notes\\
\hline
 & (this work) & (J2000) & & &  \\
\hline
- & 156  &  13.64214 -37.69327  &  -0.81 $\pm$ 0.02  &  0.94 $\pm$ 0.05 & not in \citeauthor{stasinska13}\\
14 & 420  &  13.66231 -37.66201  &  -0.72 $\pm$ 0.03  &  1.05 $\pm$ 0.07 & \\
25 & 243  &  13.68520 -37.69145  &  -0.53 $\pm$ 0.02  &  1.12 $\pm$ 0.06 & \\
35 & 566  &  13.70184 -37.66338  &  -0.47 $\pm$ 0.02  &  1.09 $\pm$ 0.05 & \\
40 & 189  &  13.71730 -37.71197  &  -1.03 $\pm$ 0.02  &  0.98 $\pm$ 0.06 & \\
45 & 653  &  13.72515 -37.65776  &  -1.03 $\pm$ 0.02  &  0.87 $\pm$ 0.05 & \\
48 & 389  &  13.72891 -37.69242  &  -0.40 $\pm$ 0.02  &  1.07 $\pm$ 0.06 & \\
51 & 403  &  13.73061 -37.69117  &  -0.88 $\pm$ 0.02  &  0.86 $\pm$ 0.04 & \\
58 & 565  &  13.74329 -37.67894  &  -0.16 $\pm$ 0.02  &  0.87 $\pm$ 0.05 & \\
63 & 735  &  13.74997 -37.65705  &  -0.96 $\pm$ 0.01  &  0.47 $\pm$ 0.04 & \\
66 & 676  &  13.76018 -37.66512  &  -0.99 $\pm$ 0.02  &  1.02 $\pm$ 0.06 & \\
69 & 622  &  13.76772 -37.68118  &  -0.84 $\pm$ 0.02  &  0.79 $\pm$ 0.05 & \\
74 & 426  &  13.77435 -37.70355  &  -1.06 $\pm$ 0.01  &  0.62 $\pm$ 0.03 & \\
\hline 
12 & - & 13.65788 -37.67055 & - & - & below detection threshold \\
20 & 374 & 13.67405 -37.67259 & -0.73$\pm$0.03 & -0.01$\pm$0.09 & \HII\ region \\
22 & - & 13.67596 -37.66800 & - & - & in \HII\ region contour \\
24 & - & 13.68208 -37.69758 & - & - & below detection threshold\\
25 & - & 13.68508 -37.69150 & - & - & below detection threshold\\
65 & - & 13.75713 -37.67483 & - & - & low signal spectrum \\
\hline
\hline
\label{tab:pne}
\end{tabular}
\end{center}
\end{table*}

\onecolumn
\begin{landscape}
\begin{center}
\begin{longtable}{lccccccccccc}
\hline 
\hline
id & Coordinates & $R$ & radius & $12+\log(\mathrm{O/H})$ & $\log([\NII]/\ha)$ & $\log([\OIII]/\hb)$ & $\log([\SII]/\ha)$ & $\log(P_\mathrm{dir}/k_\mathrm{b})$ & $\log(P_{\HII}/k_\mathrm{b})$ & $\eta_{[\NII]}$ & S32 \\
 & (J2000) & ($R_{25}$) & (pc) & & & & & (K~cm$^{-3}$) & (K~cm$^{-3}$) & & \\
\hline
9  &  13.70028   -37.72589  &  0.3  &  9.33  &  8.38 $\pm$ 0.0  &  -0.8 $\pm$ 0.01  &  -0.03 $\pm$ 0.03  &  -0.68 $\pm$ 0.01  &  4.47 $\pm$ 0.01  &  6.44 $\pm$ 0.07  &  0.12 $\pm$ 0.03  &  -0.62 $\pm$ 0.09 \\
10  &  13.6782   -37.71524  &  0.31  &  78.44  &  8.4 $\pm$ 0.0  &  -0.75 $\pm$ 0.01  &  -0.11 $\pm$ 0.04  &  -0.45 $\pm$ 0.01  &  4.41 $\pm$ 0.01  &  6.06 $\pm$ 0.04  &  0.08 $\pm$ 0.02  &  -0.94 $\pm$ 0.12 \\
11  &  13.66255   -37.70863  &  0.35  &  70.09  &  8.38 $\pm$ 0.01  &  -0.78 $\pm$ 0.01  &  -0.12 $\pm$ 0.06  &  -0.39 $\pm$ 0.01  &  3.84 $\pm$ 0.02  &  5.99 $\pm$ 0.03  &  0.08 $\pm$ 0.04  &  -0.92 $\pm$ 0.18 \\
43  &  13.72267   -37.72879  &  0.29  &  60.67  &  8.41 $\pm$ 0.0  &  -0.73 $\pm$ 0.01  &  -0.12 $\pm$ 0.03  &  -0.41 $\pm$ 0.01  &  4.51 $\pm$ 0.01  &  6.1 $\pm$ 0.04  &  0.1 $\pm$ 0.02  &  -0.84 $\pm$ 0.08 \\
47  &  13.72524   -37.73352  &  0.32  &  17.99  &  8.44 $\pm$ 0.01  &  -0.66 $\pm$ 0.02  &  -0.35 $\pm$ 0.11  &  -0.31 $\pm$ 0.02  &  4.02 $\pm$ 0.02  &  5.89 $\pm$ 0.02  &  0.15 $\pm$ 0.09  &  -0.96 $\pm$ 0.22 \\
61  &  13.74333   -37.73854  &  0.37  &  22.99  &  8.39 $\pm$ 0.01  &  -0.77 $\pm$ 0.02  &  -0.72 $\pm$ 0.21  &  -0.43 $\pm$ 0.02  &  3.99 $\pm$ 0.02  &  4.91 $\pm$ nan  &  0.22 $\pm$ 0.21  &  -1.05 $\pm$ 0.35 \\
75  &  13.71703   -37.72884  &  0.29  &  12.24  &  8.43 $\pm$ 0.01  &  -0.68 $\pm$ 0.02  &  -0.36 $\pm$ 0.13  &  -0.3 $\pm$ 0.02  &  3.78 $\pm$ 0.03  &  6.03 $\pm$ 0.04  &  0.11 $\pm$ 0.09  &  -1.08 $\pm$ 0.31 \\
99  &  13.74963   -37.73783  &  0.38  &  25.31  &  8.36 $\pm$ 0.0  &  -0.84 $\pm$ 0.01  &  0.2 $\pm$ 0.02  &  -0.55 $\pm$ 0.01  &  4.65 $\pm$ 0.01  &  5.98 $\pm$ 0.03  &  0.07 $\pm$ 0.01  &  -0.61 $\pm$ 0.08 \\
101  &  13.73903   -37.73425  &  0.34  &  18.29  &  8.42 $\pm$ 0.01  &  -0.7 $\pm$ 0.01  &  -0.48 $\pm$ 0.1  &  -0.32 $\pm$ 0.01  &  4.12 $\pm$ 0.02  &  5.9 $\pm$ 0.02  &  0.12 $\pm$ 0.08  &  -1.15 $\pm$ 0.26 \\
121  &  13.75728   -37.73758  &  0.39  &  38.49  &  8.35 $\pm$ 0.0  &  -0.84 $\pm$ 0.01  &  0.24 $\pm$ 0.02  &  -0.61 $\pm$ 0.01  &  4.49 $\pm$ 0.01  &  5.85 $\pm$ 0.01  &  0.06 $\pm$ 0.01  &  -0.57 $\pm$ 0.09 \\
129  &  13.74591   -37.73335  &  0.34  &  26.38  &  8.4 $\pm$ 0.0  &  -0.75 $\pm$ 0.01  &  0.4 $\pm$ 0.02  &  -0.46 $\pm$ 0.01  &  4.51 $\pm$ 0.01  &  5.56 $\pm$ nan  &  0.04 $\pm$ 0.01  &  -0.69 $\pm$ 0.09 \\
131  &  13.71105   -37.72053  &  0.24  &  21.24  &  8.41 $\pm$ 0.0  &  -0.73 $\pm$ 0.01  &  -0.22 $\pm$ 0.05  &  -0.52 $\pm$ 0.01  &  4.3 $\pm$ 0.01  &  6.18 $\pm$ 0.05  &  0.2 $\pm$ 0.06  &  -0.64 $\pm$ 0.1 \\
143  &  13.75131   -37.73545  &  0.37  &  10.36  &  8.43 $\pm$ 0.01  &  -0.67 $\pm$ 0.02  &  -0.0 $\pm$ 0.07  &  -0.32 $\pm$ 0.02  &  3.94 $\pm$ 0.03  &  4.91 $\pm$ nan  &  0.08 $\pm$ 0.04  &  -0.9 $\pm$ 0.23 \\
153  &  13.69896   -37.71291  &  0.22  &  26.78  &  8.36 $\pm$ 0.01  &  -0.84 $\pm$ 0.02  &  0.54 $\pm$ 0.05  &  -0.58 $\pm$ 0.02  &  4.46 $\pm$ 0.02  &  6.24 $\pm$ 0.08  &  0.01 $\pm$ 0.01  &  -0.96 $\pm$ 0.18 \\
168  &  13.6601   -37.69548  &  0.33  &  23.7  &  8.39 $\pm$ 0.0  &  -0.76 $\pm$ 0.0  &  0.07 $\pm$ 0.01  &  -0.61 $\pm$ 0.0  &  4.97 $\pm$ 0.0  &  6.26 $\pm$ 0.05  &  0.12 $\pm$ 0.01  &  -0.56 $\pm$ 0.03 \\
173  &  13.74078   -37.72636  &  0.29  &  9.1  &  8.39 $\pm$ 0.01  &  -0.75 $\pm$ 0.01  &  -0.2 $\pm$ 0.06  &  -0.59 $\pm$ 0.01  &  4.24 $\pm$ 0.02  &  5.3 $\pm$ nan  &  0.2 $\pm$ 0.08  &  -0.61 $\pm$ 0.16 \\
179  &  13.66797   -37.69777  &  0.3  &  12.13  &  8.36 $\pm$ 0.01  &  -0.82 $\pm$ 0.02  &  -0.72 $\pm$ 0.22  &  -0.39 $\pm$ 0.02  &  4.04 $\pm$ 0.02  &  6.04 $\pm$ 0.04  &  0.16 $\pm$ 0.16  &  -1.15 $\pm$ 0.35 \\
180  &  13.66301   -37.69568  &  0.32  &  7.09  &  8.41 $\pm$ 0.0  &  -0.71 $\pm$ 0.01  &  -0.48 $\pm$ 0.09  &  -0.49 $\pm$ 0.01  &  4.34 $\pm$ 0.01  &  6.16 $\pm$ 0.05  &  0.28 $\pm$ 0.1  &  -0.78 $\pm$ 0.12 \\
187  &  13.66121   -37.69063  &  0.32  &  7.67  &  8.41 $\pm$ 0.0  &  -0.72 $\pm$ 0.0  &  -0.21 $\pm$ 0.02  &  -0.66 $\pm$ 0.0  &  5.11 $\pm$ 0.0  &  6.26 $\pm$ 0.05  &  0.3 $\pm$ 0.02  &  -0.48 $\pm$ 0.03 \\
193  &  13.6858   -37.6988  &  0.21  &  10.27  &  8.43 $\pm$ 0.0  &  -0.67 $\pm$ 0.01  &  -0.56 $\pm$ 0.08  &  -0.48 $\pm$ 0.01  &  4.39 $\pm$ 0.01  &  5.99 $\pm$ 0.03  &  0.4 $\pm$ 0.12  &  -0.74 $\pm$ 0.1 \\
195  &  13.64618   -37.68315  &  0.4  &  7.64  &  8.42 $\pm$ 0.01  &  -0.7 $\pm$ 0.02  &  -0.78 $\pm$ 0.24  &  -0.46 $\pm$ 0.02  &  4.06 $\pm$ 0.02  &  6.25 $\pm$ 0.06  &  0.57 $\pm$ 0.4  &  -0.79 $\pm$ 0.18 \\
198  &  13.65941   -37.68699  &  0.33  &  14.42  &  8.44 $\pm$ 0.01  &  -0.66 $\pm$ 0.02  &  -0.75 $\pm$ 0.22  &  -0.22 $\pm$ 0.02  &  4.02 $\pm$ 0.02  &  6.16 $\pm$ 0.05  &  0.25 $\pm$ 0.19  &  -1.16 $\pm$ 0.23 \\
200  &  13.68818   -37.69692  &  0.2  &  19.04  &  8.41 $\pm$ 0.0  &  -0.72 $\pm$ 0.0  &  -0.35 $\pm$ 0.03  &  -0.57 $\pm$ 0.0  &  4.78 $\pm$ 0.01  &  6.37 $\pm$ 0.05  &  0.32 $\pm$ 0.04  &  -0.59 $\pm$ 0.04 \\
201  &  13.6484   -37.68201  &  0.38  &  13.19  &  8.41 $\pm$ 0.01  &  -0.73 $\pm$ 0.02  &  -0.56 $\pm$ 0.17  &  -0.38 $\pm$ 0.02  &  3.99 $\pm$ 0.02  &  6.27 $\pm$ 0.06  &  0.23 $\pm$ 0.15  &  -0.92 $\pm$ 0.21 \\
205  &  13.65381   -37.68245  &  0.36  &  32.14  &  8.48 $\pm$ 0.01  &  -0.57 $\pm$ 0.02  &  0.42 $\pm$ 0.05  &  -0.3 $\pm$ 0.02  &  3.73 $\pm$ 0.02  &  6.22 $\pm$ 0.05  &  0.05 $\pm$ 0.02  &  -0.75 $\pm$ 0.12 \\
211  &  13.75503   -37.72081  &  0.29  &  15.97  &  8.41 $\pm$ 0.01  &  -0.73 $\pm$ 0.02  &  -0.65 $\pm$ 0.25  &  -0.33 $\pm$ 0.02  &  4.0 $\pm$ 0.03  &  6.35 $\pm$ 0.07  &  0.17 $\pm$ 0.17  &  -1.15 $\pm$ 0.34 \\
216  &  13.76553   -37.72171  &  0.33  &  56.7  &  8.4 $\pm$ 0.0  &  -0.74 $\pm$ 0.0  &  -0.08 $\pm$ 0.01  &  -0.57 $\pm$ 0.0  &  5.02 $\pm$ 0.0  &  6.26 $\pm$ 0.04  &  0.11 $\pm$ 0.01  &  -0.75 $\pm$ 0.03 \\
221  &  13.66576   -37.68678  &  0.29  &  8.76  &  8.37 $\pm$ 0.01  &  -0.81 $\pm$ 0.02  &  -0.37 $\pm$ 0.15  &  -0.35 $\pm$ 0.02  &  3.77 $\pm$ 0.03  &  5.89 $\pm$ 0.02  &  0.09 $\pm$ 0.08  &  -1.07 $\pm$ 0.37 \\
240  &  13.72264   -37.70547  &  0.14  &  9.13  &  8.44 $\pm$ 0.0  &  -0.65 $\pm$ 0.01  &  -0.88 $\pm$ 0.13  &  -0.49 $\pm$ 0.01  &  4.58 $\pm$ 0.01  &  6.02 $\pm$ 0.03  &  0.77 $\pm$ 0.27  &  -0.8 $\pm$ 0.09 \\
241  &  13.6686   -37.68198  &  0.28  &  35.07  &  8.44 $\pm$ 0.0  &  -0.65 $\pm$ 0.0  &  -0.31 $\pm$ 0.02  &  -0.4 $\pm$ 0.0  &  4.95 $\pm$ 0.01  &  6.19 $\pm$ 0.04  &  0.17 $\pm$ 0.02  &  -0.89 $\pm$ 0.04 \\
249  &  13.64161   -37.6714  &  0.43  &  27.6  &  8.43 $\pm$ 0.01  &  -0.68 $\pm$ 0.01  &  0.33 $\pm$ 0.04  &  -0.38 $\pm$ 0.01  &  4.07 $\pm$ 0.02  &  6.23 $\pm$ 0.06  &  0.05 $\pm$ 0.02  &  -0.73 $\pm$ 0.11 \\
253  &  13.66312   -37.68058  &  0.31  &  14.97  &  8.42 $\pm$ 0.0  &  -0.69 $\pm$ 0.01  &  -0.35 $\pm$ 0.07  &  -0.5 $\pm$ 0.01  &  4.25 $\pm$ 0.01  &  6.15 $\pm$ 0.05  &  0.32 $\pm$ 0.09  &  -0.62 $\pm$ 0.09 \\
259  &  13.65466   -37.6774  &  0.35  &  10.81  &  8.44 $\pm$ 0.0  &  -0.65 $\pm$ 0.01  &  -0.62 $\pm$ 0.1  &  -0.51 $\pm$ 0.01  &  4.39 $\pm$ 0.01  &  6.14 $\pm$ 0.04  &  0.53 $\pm$ 0.17  &  -0.7 $\pm$ 0.09 \\
264  &  13.74043   -37.70672  &  0.17  &  33.39  &  8.43 $\pm$ 0.0  &  -0.68 $\pm$ 0.01  &  0.02 $\pm$ 0.03  &  -0.58 $\pm$ 0.01  &  4.38 $\pm$ 0.01  &  6.12 $\pm$ 0.04  &  0.18 $\pm$ 0.04  &  -0.5 $\pm$ 0.07 \\
271  &  13.73084   -37.70376  &  0.13  &  27.13  &  8.43 $\pm$ 0.0  &  -0.67 $\pm$ 0.01  &  -0.79 $\pm$ 0.14  &  -0.46 $\pm$ 0.01  &  4.36 $\pm$ 0.01  &  6.1 $\pm$ 0.04  &  0.51 $\pm$ 0.23  &  -0.87 $\pm$ 0.14 \\
273  &  13.70943   -37.69686  &  0.11  &  14.36  &  8.43 $\pm$ 0.01  &  -0.68 $\pm$ 0.02  &  -0.76 $\pm$ 0.25  &  -0.31 $\pm$ 0.02  &  3.86 $\pm$ 0.02  &  5.66 $\pm$ nan  &  0.24 $\pm$ 0.25  &  -1.16 $\pm$ 0.35 \\
276  &  13.68677   -37.68616  &  0.19  &  27.2  &  8.43 $\pm$ 0.01  &  -0.68 $\pm$ 0.01  &  -0.9 $\pm$ 0.23  &  -0.42 $\pm$ 0.01  &  4.16 $\pm$ 0.02  &  5.87 $\pm$ 0.01  &  0.41 $\pm$ 0.32  &  -1.06 $\pm$ 0.23 \\
277  &  13.71849   -37.69948  &  0.1  &  12.58  &  8.44 $\pm$ 0.01  &  -0.66 $\pm$ 0.01  &  0.16 $\pm$ 0.05  &  -0.46 $\pm$ 0.01  &  4.07 $\pm$ 0.02  &  6.1 $\pm$ 0.05  &  0.08 $\pm$ 0.03  &  -0.73 $\pm$ 0.17 \\
278  &  13.74575   -37.7098  &  0.21  &  9.08  &  8.45 $\pm$ 0.01  &  -0.63 $\pm$ 0.01  &  -0.69 $\pm$ 0.17  &  -0.44 $\pm$ 0.01  &  4.16 $\pm$ 0.02  &  6.17 $\pm$ 0.05  &  0.51 $\pm$ 0.3  &  -0.8 $\pm$ 0.18 \\
280  &  13.76577   -37.71332  &  0.29  &  55.79  &  8.39 $\pm$ 0.0  &  -0.75 $\pm$ 0.0  &  0.07 $\pm$ 0.01  &  -0.49 $\pm$ 0.0  &  4.8 $\pm$ 0.0  &  6.13 $\pm$ 0.04  &  0.07 $\pm$ 0.01  &  -0.79 $\pm$ 0.04 \\
286  &  13.71583   -37.69476  &  0.08  &  36.58  &  8.45 $\pm$ 0.0  &  -0.63 $\pm$ 0.01  &  -0.63 $\pm$ 0.09  &  -0.42 $\pm$ 0.01  &  4.28 $\pm$ 0.01  &  6.13 $\pm$ 0.04  &  0.2 $\pm$ 0.11  &  -1.16 $\pm$ 0.21 \\
292  &  13.72736   -37.7016  &  0.12  &  7.81  &  8.47 $\pm$ 0.01  &  -0.58 $\pm$ 0.01  &  -1.0 $\pm$ 0.27  &  -0.4 $\pm$ 0.01  &  4.24 $\pm$ 0.02  &  6.06 $\pm$ 0.04  &  0.83 $\pm$ 0.63  &  -0.95 $\pm$ 0.18 \\
295  &  13.70713   -37.69345  &  0.1  &  11.58  &  8.46 $\pm$ 0.01  &  -0.62 $\pm$ 0.02  &  -0.87 $\pm$ 0.27  &  -0.39 $\pm$ 0.02  &  4.03 $\pm$ 0.02  &  5.9 $\pm$ 0.02  &  0.46 $\pm$ 0.42  &  -1.04 $\pm$ 0.28 \\
309  &  13.70391   -37.69022  &  0.1  &  14.51  &  8.47 $\pm$ 0.01  &  -0.59 $\pm$ 0.01  &  -0.59 $\pm$ 0.12  &  -0.41 $\pm$ 0.01  &  4.22 $\pm$ 0.02  &  5.98 $\pm$ 0.03  &  0.33 $\pm$ 0.16  &  -0.94 $\pm$ 0.17 \\
326  &  13.72789   -37.69846  &  0.1  &  23.81  &  8.43 $\pm$ 0.01  &  -0.68 $\pm$ 0.02  &  -0.92 $\pm$ 0.31  &  -0.29 $\pm$ 0.02  &  3.77 $\pm$ 0.02  &  6.03 $\pm$ 0.04  &  0.37 $\pm$ 0.39  &  -1.13 $\pm$ 0.33 \\
328  &  13.77346   -37.71463  &  0.33  &  24.97  &  8.45 $\pm$ 0.0  &  -0.64 $\pm$ 0.01  &  -0.28 $\pm$ 0.04  &  -0.38 $\pm$ 0.01  &  4.12 $\pm$ 0.01  &  6.07 $\pm$ 0.03  &  0.15 $\pm$ 0.04  &  -0.91 $\pm$ 0.1 \\
330  &  13.67193   -37.67777  &  0.27  &  8.23  &  8.44 $\pm$ 0.01  &  -0.66 $\pm$ 0.01  &  -0.71 $\pm$ 0.17  &  -0.42 $\pm$ 0.01  &  4.11 $\pm$ 0.02  &  6.24 $\pm$ 0.06  &  0.41 $\pm$ 0.24  &  -0.89 $\pm$ 0.18 \\
345  &  13.74381   -37.70302  &  0.16  &  9.7  &  8.33 $\pm$ 0.01  &  -0.88 $\pm$ 0.01  &  -1.21 $\pm$ 0.33  &  -0.47 $\pm$ 0.01  &  4.1 $\pm$ 0.02  &  5.34 $\pm$ nan  &  0.34 $\pm$ 0.49  &  -1.25 $\pm$ 0.52 \\
348  &  13.71003   -37.68504  &  0.07  &  62.28  &  8.46 $\pm$ 0.0  &  -0.62 $\pm$ 0.01  &  -0.64 $\pm$ 0.13  &  -0.44 $\pm$ 0.01  &  3.93 $\pm$ 0.02  &  5.97 $\pm$ 0.03  &  0.34 $\pm$ 0.17  &  -0.95 $\pm$ 0.18 \\
359  &  13.64988   -37.66397  &  0.4  &  23.09  &  8.39 $\pm$ 0.0  &  -0.77 $\pm$ 0.01  &  -0.05 $\pm$ 0.03  &  -0.64 $\pm$ 0.01  &  4.55 $\pm$ 0.01  &  6.31 $\pm$ 0.06  &  0.18 $\pm$ 0.03  &  -0.48 $\pm$ 0.06 \\
361  &  13.67911   -37.66773  &  0.25  &  56.54  &  8.43 $\pm$ 0.0  &  -0.68 $\pm$ 0.01  &  -0.03 $\pm$ 0.04  &  -0.43 $\pm$ 0.01  &  4.22 $\pm$ 0.01  &  6.2 $\pm$ 0.05  &  0.11 $\pm$ 0.03  &  -0.76 $\pm$ 0.1 \\
374  &  13.67405   -37.67259  &  0.26  &  11.18  &  8.41 $\pm$ 0.01  &  -0.73 $\pm$ 0.02  &  -0.01 $\pm$ 0.09  &  -0.46 $\pm$ 0.02  &  3.78 $\pm$ 0.03  &  6.55 $\pm$ 0.12  &  0.15 $\pm$ 0.08  &  -0.57 $\pm$ 0.17 \\
375  &  13.6842   -37.67406  &  0.21  &  21.86  &  8.43 $\pm$ 0.0  &  -0.68 $\pm$ 0.0  &  -0.25 $\pm$ 0.02  &  -0.62 $\pm$ 0.0  &  4.82 $\pm$ 0.01  &  6.06 $\pm$ 0.03  &  0.3 $\pm$ 0.04  &  -0.55 $\pm$ 0.04 \\
378  &  13.64784   -37.66067  &  0.42  &  24.09  &  8.34 $\pm$ 0.0  &  -0.86 $\pm$ 0.01  &  -0.0 $\pm$ 0.03  &  -0.58 $\pm$ 0.01  &  4.28 $\pm$ 0.01  &  6.29 $\pm$ 0.06  &  0.06 $\pm$ 0.02  &  -0.83 $\pm$ 0.11 \\
386  &  13.76038   -37.70391  &  0.23  &  9.09  &  8.45 $\pm$ 0.0  &  -0.63 $\pm$ 0.01  &  -0.56 $\pm$ 0.1  &  -0.4 $\pm$ 0.01  &  4.27 $\pm$ 0.01  &  5.67 $\pm$ nan  &  0.29 $\pm$ 0.15  &  -0.92 $\pm$ 0.21 \\
412  &  13.69381   -37.67284  &  0.17  &  26.69  &  8.47 $\pm$ 0.0  &  -0.59 $\pm$ 0.01  &  -0.65 $\pm$ 0.1  &  -0.41 $\pm$ 0.01  &  4.38 $\pm$ 0.01  &  5.85 $\pm$ 0.01  &  0.37 $\pm$ 0.14  &  -0.95 $\pm$ 0.13 \\
418  &  13.67502   -37.66542  &  0.28  &  18.75  &  8.45 $\pm$ 0.01  &  -0.63 $\pm$ 0.02  &  -0.38 $\pm$ 0.16  &  -0.27 $\pm$ 0.02  &  3.51 $\pm$ 0.03  &  6.21 $\pm$ 0.06  &  0.16 $\pm$ 0.12  &  -1.01 $\pm$ 0.26 \\
421  &  13.75274   -37.69474  &  0.17  &  23.08  &  8.47 $\pm$ 0.0  &  -0.59 $\pm$ 0.01  &  -0.33 $\pm$ 0.07  &  -0.52 $\pm$ 0.01  &  4.2 $\pm$ 0.01  &  5.96 $\pm$ 0.03  &  0.37 $\pm$ 0.12  &  -0.63 $\pm$ 0.12 \\
423  &  13.73544   -37.68682  &  0.07  &  40.16  &  8.48 $\pm$ 0.0  &  -0.56 $\pm$ 0.01  &  -0.55 $\pm$ 0.07  &  -0.4 $\pm$ 0.01  &  4.44 $\pm$ 0.01  &  6.17 $\pm$ 0.04  &  0.39 $\pm$ 0.12  &  -0.85 $\pm$ 0.11 \\
433  &  13.71171   -37.67379  &  0.09  &  80.19  &  8.47 $\pm$ 0.0  &  -0.58 $\pm$ 0.01  &  -0.4 $\pm$ 0.05  &  -0.39 $\pm$ 0.01  &  4.48 $\pm$ 0.01  &  6.03 $\pm$ 0.03  &  0.24 $\pm$ 0.06  &  -0.9 $\pm$ 0.09 \\
440  &  13.76278   -37.69634  &  0.22  &  30.77  &  8.47 $\pm$ 0.01  &  -0.59 $\pm$ 0.01  &  -0.88 $\pm$ 0.27  &  -0.29 $\pm$ 0.01  &  4.05 $\pm$ 0.02  &  5.84 $\pm$ 0.01  &  0.34 $\pm$ 0.32  &  -1.23 $\pm$ 0.31 \\
443  &  13.68593   -37.66689  &  0.22  &  25.63  &  8.43 $\pm$ 0.0  &  -0.68 $\pm$ 0.01  &  -0.31 $\pm$ 0.05  &  -0.5 $\pm$ 0.01  &  4.4 $\pm$ 0.01  &  6.16 $\pm$ 0.05  &  0.23 $\pm$ 0.06  &  -0.73 $\pm$ 0.09 \\
469  &  13.69056   -37.66822  &  0.2  &  17.59  &  8.44 $\pm$ 0.01  &  -0.66 $\pm$ 0.02  &  -0.13 $\pm$ 0.08  &  -0.32 $\pm$ 0.02  &  3.91 $\pm$ 0.03  &  5.9 $\pm$ 0.02  &  0.1 $\pm$ 0.05  &  -0.94 $\pm$ 0.22 \\
489  &  13.74339   -37.68762  &  0.11  &  7.32  &  8.46 $\pm$ 0.01  &  -0.61 $\pm$ 0.02  &  0.55 $\pm$ 0.05  &  -0.27 $\pm$ 0.02  &  3.97 $\pm$ 0.02  &  6.23 $\pm$ 0.06  &  0.02 $\pm$ 0.01  &  -0.96 $\pm$ 0.21 \\
509  &  13.72713   -37.67955  &  0.04  &  8.08  &  8.5 $\pm$ 0.01  &  -0.53 $\pm$ 0.01  &  -0.62 $\pm$ 0.14  &  -0.38 $\pm$ 0.01  &  4.2 $\pm$ 0.02  &  5.63 $\pm$ nan  &  0.35 $\pm$ 0.2  &  -1.0 $\pm$ 0.21 \\
512  &  13.74734   -37.68616  &  0.13  &  16.81  &  8.52 $\pm$ 0.01  &  -0.49 $\pm$ 0.02  &  -0.44 $\pm$ 0.12  &  -0.17 $\pm$ 0.02  &  3.96 $\pm$ 0.02  &  5.55 $\pm$ nan  &  0.15 $\pm$ 0.11  &  -1.22 $\pm$ 0.27 \\
513  &  13.74012   -37.68331  &  0.09  &  13.41  &  8.53 $\pm$ 0.0  &  -0.46 $\pm$ 0.01  &  0.45 $\pm$ 0.02  &  -0.31 $\pm$ 0.01  &  4.5 $\pm$ 0.01  &  5.87 $\pm$ 0.01  &  0.07 $\pm$ 0.01  &  -0.7 $\pm$ 0.06 \\
514  &  13.78151   -37.69698  &  0.32  &  35.64  &  8.35 $\pm$ 0.01  &  -0.85 $\pm$ 0.03  &  0.33 $\pm$ 0.09  &  -0.64 $\pm$ 0.03  &  4.56 $\pm$ 0.04  &  6.08 $\pm$ 0.11  &  0.05 $\pm$ 0.03  &  -0.57 $\pm$ 0.13 \\
523  &  13.77118   -37.69105  &  0.26  &  44.32  &  8.45 $\pm$ 0.02  &  -0.63 $\pm$ 0.05  &  -0.07 $\pm$ 0.23  &  -0.34 $\pm$ 0.05  &  4.22 $\pm$ 0.08  &  5.99 $\pm$ 0.07  &  0.08 $\pm$ 0.08  &  -0.97 $\pm$ 0.29 \\
536  &  13.73535   -37.67277  &  0.1  &  80.79  &  8.46 $\pm$ 0.01  &  -0.61 $\pm$ 0.01  &  0.16 $\pm$ 0.04  &  -0.42 $\pm$ 0.01  &  4.04 $\pm$ 0.02  &  5.41 $\pm$ nan  &  0.11 $\pm$ 0.03  &  -0.65 $\pm$ 0.11 \\
545  &  13.73902   -37.67861  &  0.09  &  11.2  &  8.52 $\pm$ 0.01  &  -0.47 $\pm$ 0.02  &  0.32 $\pm$ 0.06  &  -0.3 $\pm$ 0.02  &  4.02 $\pm$ 0.03  &  nan $\pm$ nan  &  0.13 $\pm$ 0.03  &  -0.56 $\pm$ 0.1 \\
551  &  13.70873   -37.66656  &  0.14  &  12.54  &  8.45 $\pm$ 0.01  &  -0.63 $\pm$ 0.02  &  -0.7 $\pm$ 0.2  &  -0.29 $\pm$ 0.02  &  4.05 $\pm$ 0.02  &  5.45 $\pm$ nan  &  0.26 $\pm$ 0.19  &  -1.11 $\pm$ 0.24 \\
554  &  13.68307   -37.65676  &  0.27  &  9.87  &  8.44 $\pm$ 0.0  &  -0.66 $\pm$ 0.01  &  -0.23 $\pm$ 0.05  &  -0.5 $\pm$ 0.01  &  4.46 $\pm$ 0.01  &  5.96 $\pm$ 0.03  &  0.23 $\pm$ 0.05  &  -0.68 $\pm$ 0.08 \\
556  &  13.75273   -37.6819  &  0.16  &  25.3  &  8.47 $\pm$ 0.01  &  -0.58 $\pm$ 0.01  &  -0.53 $\pm$ 0.13  &  -0.37 $\pm$ 0.01  &  4.14 $\pm$ 0.02  &  nan $\pm$ nan  &  0.3 $\pm$ 0.16  &  -0.92 $\pm$ 0.19 \\
560  &  13.67659   -37.65052  &  0.33  &  41.47  &  8.4 $\pm$ 0.0  &  -0.73 $\pm$ 0.01  &  -0.29 $\pm$ 0.04  &  -0.56 $\pm$ 0.01  &  4.46 $\pm$ 0.01  &  5.78 $\pm$ nan  &  0.23 $\pm$ 0.05  &  -0.64 $\pm$ 0.07 \\
564  &  13.76782   -37.68801  &  0.24  &  8.48  &  8.46 $\pm$ 0.04  &  -0.61 $\pm$ 0.09  &  -0.35 $\pm$ 0.63  &  -0.38 $\pm$ 0.09  &  3.99 $\pm$ 0.13  &  6.06 $\pm$ 0.18  &  0.17 $\pm$ 0.34  &  -0.97 $\pm$ 0.52 \\
572  &  13.75247   -37.67682  &  0.16  &  34.61  &  8.45 $\pm$ 0.0  &  -0.64 $\pm$ 0.0  &  -0.33 $\pm$ 0.03  &  -0.5 $\pm$ 0.0  &  4.75 $\pm$ 0.01  &  5.9 $\pm$ 0.02  &  0.28 $\pm$ 0.04  &  -0.71 $\pm$ 0.06 \\
581  &  13.71629   -37.66566  &  0.13  &  10.66  &  8.5 $\pm$ 0.01  &  -0.52 $\pm$ 0.02  &  -0.35 $\pm$ 0.13  &  -0.13 $\pm$ 0.02  &  3.73 $\pm$ 0.03  &  5.52 $\pm$ nan  &  0.17 $\pm$ 0.1  &  -1.05 $\pm$ 0.21 \\
593  &  13.74934   -37.67476  &  0.15  &  18.33  &  8.48 $\pm$ 0.01  &  -0.56 $\pm$ 0.02  &  -0.48 $\pm$ 0.15  &  -0.26 $\pm$ 0.02  &  3.93 $\pm$ 0.03  &  nan $\pm$ nan  &  0.27 $\pm$ 0.17  &  -0.94 $\pm$ 0.22 \\
601  &  13.70513   -37.65722  &  0.2  &  34.49  &  8.4 $\pm$ 0.0  &  -0.75 $\pm$ 0.01  &  -0.82 $\pm$ 0.15  &  -0.45 $\pm$ 0.01  &  4.32 $\pm$ 0.01  &  5.51 $\pm$ nan  &  0.37 $\pm$ 0.18  &  -0.95 $\pm$ 0.14 \\
602  &  13.76977   -37.68269  &  0.24  &  29.49  &  8.45 $\pm$ 0.01  &  -0.63 $\pm$ 0.02  &  -0.14 $\pm$ 0.08  &  -0.55 $\pm$ 0.02  &  4.65 $\pm$ 0.02  &  6.1 $\pm$ 0.06  &  0.23 $\pm$ 0.06  &  -0.6 $\pm$ 0.07 \\
606  &  13.6748   -37.64391  &  0.36  &  51.55  &  8.39 $\pm$ 0.01  &  -0.76 $\pm$ 0.01  &  0.03 $\pm$ 0.05  &  -0.52 $\pm$ 0.01  &  4.08 $\pm$ 0.02  &  6.1 $\pm$ 0.05  &  0.13 $\pm$ 0.04  &  -0.53 $\pm$ 0.09 \\
613  &  13.78064   -37.68524  &  0.3  &  25.32  &  8.42 $\pm$ 0.01  &  -0.71 $\pm$ 0.03  &  -0.27 $\pm$ 0.19  &  -0.57 $\pm$ 0.03  &  4.64 $\pm$ 0.05  &  5.86 $\pm$ 0.03  &  0.23 $\pm$ 0.14  &  -0.65 $\pm$ 0.13 \\
620  &  13.71649   -37.6607  &  0.16  &  18.11  &  8.33 $\pm$ 0.0  &  -0.9 $\pm$ 0.0  &  0.31 $\pm$ 0.01  &  -0.92 $\pm$ 0.0  &  5.62 $\pm$ 0.0  &  6.31 $\pm$ 0.06  &  0.13 $\pm$ 0.01  &  -0.13 $\pm$ 0.01 \\
630  &  13.6989   -37.65402  &  0.23  &  9.26  &  8.45 $\pm$ 0.01  &  -0.63 $\pm$ 0.02  &  -0.5 $\pm$ 0.2  &  -0.29 $\pm$ 0.02  &  3.75 $\pm$ 0.03  &  nan $\pm$ nan  &  0.23 $\pm$ 0.19  &  -0.96 $\pm$ 0.27 \\
641  &  13.68863   -37.64609  &  0.31  &  26.54  &  8.35 $\pm$ 0.0  &  -0.86 $\pm$ 0.0  &  0.17 $\pm$ 0.01  &  -0.77 $\pm$ 0.0  &  5.06 $\pm$ 0.01  &  5.95 $\pm$ 0.02  &  0.1 $\pm$ 0.01  &  -0.43 $\pm$ 0.03 \\
646  &  13.70695   -37.65494  &  0.21  &  7.9  &  8.41 $\pm$ 0.01  &  -0.72 $\pm$ 0.03  &  -0.81 $\pm$ 0.26  &  -0.28 $\pm$ 0.03  &  3.7 $\pm$ 0.04  &  5.85 $\pm$ 0.01  &  0.28 $\pm$ 0.3  &  -1.1 $\pm$ 0.37 \\
649  &  13.69936   -37.64915  &  0.26  &  28.63  &  8.43 $\pm$ 0.01  &  -0.69 $\pm$ 0.02  &  -0.68 $\pm$ 0.25  &  -0.28 $\pm$ 0.02  &  3.69 $\pm$ 0.03  &  5.65 $\pm$ nan  &  0.26 $\pm$ 0.22  &  -1.03 $\pm$ 0.26 \\
654  &  13.78761   -37.68073  &  0.34  &  11.04  &  8.37 $\pm$ 0.0  &  -0.8 $\pm$ 0.01  &  0.07 $\pm$ 0.02  &  -0.67 $\pm$ 0.01  &  4.79 $\pm$ 0.01  &  5.85 $\pm$ 0.01  &  0.13 $\pm$ 0.02  &  -0.46 $\pm$ 0.05 \\
655  &  13.71162   -37.64137  &  0.29  &  85.68  &  8.38 $\pm$ 0.0  &  -0.79 $\pm$ 0.01  &  -0.07 $\pm$ 0.02  &  -0.51 $\pm$ 0.01  &  4.67 $\pm$ 0.01  &  6.03 $\pm$ 0.03  &  0.1 $\pm$ 0.02  &  -0.75 $\pm$ 0.06 \\
660  &  13.7697   -37.67247  &  0.26  &  8.83  &  8.41 $\pm$ 0.01  &  -0.73 $\pm$ 0.01  &  0.09 $\pm$ 0.05  &  -0.36 $\pm$ 0.01  &  4.11 $\pm$ 0.02  &  nan $\pm$ nan  &  0.06 $\pm$ 0.03  &  -0.84 $\pm$ 0.18 \\
667  &  13.78192   -37.67471  &  0.31  &  9.93  &  8.39 $\pm$ 0.01  &  -0.75 $\pm$ 0.01  &  0.18 $\pm$ 0.05  &  -0.38 $\pm$ 0.01  &  4.11 $\pm$ 0.02  &  nan $\pm$ nan  &  0.06 $\pm$ 0.02  &  -0.75 $\pm$ 0.16 \\
682  &  13.79612   -37.67812  &  0.38  &  8.48  &  8.43 $\pm$ 0.01  &  -0.69 $\pm$ 0.02  &  -0.56 $\pm$ 0.15  &  -0.39 $\pm$ 0.02  &  3.97 $\pm$ 0.02  &  6.02 $\pm$ 0.04  &  0.25 $\pm$ 0.19  &  -0.93 $\pm$ 0.28 \\
693  &  13.75948   -37.66241  &  0.24  &  19.54  &  8.41 $\pm$ 0.0  &  -0.72 $\pm$ 0.01  &  -0.02 $\pm$ 0.03  &  -0.54 $\pm$ 0.01  &  4.44 $\pm$ 0.01  &  5.62 $\pm$ nan  &  0.14 $\pm$ 0.03  &  -0.6 $\pm$ 0.08 \\
723  &  13.69651   -37.63373  &  0.36  &  44.31  &  8.41 $\pm$ 0.0  &  -0.71 $\pm$ 0.01  &  -0.17 $\pm$ 0.03  &  -0.29 $\pm$ 0.01  &  4.44 $\pm$ 0.01  &  6.1 $\pm$ 0.03  &  0.06 $\pm$ 0.01  &  -1.11 $\pm$ 0.09 \\
724  &  13.75731   -37.65999  &  0.24  &  12.64  &  8.46 $\pm$ 0.01  &  -0.62 $\pm$ 0.02  &  -0.27 $\pm$ 0.1  &  -0.32 $\pm$ 0.02  &  4.03 $\pm$ 0.03  &  nan $\pm$ nan  &  0.15 $\pm$ 0.08  &  -0.93 $\pm$ 0.22 \\
729  &  13.75225   -37.65732  &  0.23  &  15.58  &  8.42 $\pm$ 0.01  &  -0.71 $\pm$ 0.01  &  -0.73 $\pm$ 0.18  &  -0.43 $\pm$ 0.01  &  4.07 $\pm$ 0.02  &  nan $\pm$ nan  &  0.3 $\pm$ 0.21  &  -1.0 $\pm$ 0.24 \\
734  &  13.75614   -37.65825  &  0.24  &  21.03  &  8.44 $\pm$ 0.01  &  -0.65 $\pm$ 0.02  &  -0.51 $\pm$ 0.16  &  -0.34 $\pm$ 0.02  &  3.88 $\pm$ 0.02  &  nan $\pm$ nan  &  0.2 $\pm$ 0.14  &  -1.03 $\pm$ 0.26 \\
740  &  13.75977   -37.66005  &  0.25  &  10.68  &  8.48 $\pm$ 0.01  &  -0.56 $\pm$ 0.02  &  -0.47 $\pm$ 0.16  &  -0.28 $\pm$ 0.02  &  3.92 $\pm$ 0.03  &  nan $\pm$ nan  &  0.29 $\pm$ 0.19  &  -0.91 $\pm$ 0.22 \\
748  &  13.72862   -37.64647  &  0.25  &  20.46  &  8.44 $\pm$ 0.01  &  -0.67 $\pm$ 0.01  &  -0.61 $\pm$ 0.14  &  -0.42 $\pm$ 0.01  &  4.05 $\pm$ 0.02  &  5.81 $\pm$ 0.01  &  0.37 $\pm$ 0.19  &  -0.83 $\pm$ 0.17 \\
751  &  13.68959   -37.63199  &  0.38  &  18.29  &  8.37 $\pm$ 0.0  &  -0.81 $\pm$ 0.0  &  -0.04 $\pm$ 0.01  &  -0.69 $\pm$ 0.0  &  5.06 $\pm$ 0.0  &  6.08 $\pm$ 0.03  &  0.15 $\pm$ 0.01  &  -0.51 $\pm$ 0.03 \\
752  &  13.72298   -37.64481  &  0.26  &  16.18  &  8.46 $\pm$ 0.01  &  -0.62 $\pm$ 0.02  &  -0.09 $\pm$ 0.07  &  -0.46 $\pm$ 0.02  &  3.98 $\pm$ 0.02  &  5.44 $\pm$ nan  &  0.21 $\pm$ 0.08  &  -0.6 $\pm$ 0.13 \\
790  &  13.70286   -37.6348  &  0.34  &  7.37  &  8.31 $\pm$ 0.01  &  -0.94 $\pm$ 0.02  &  -1.02 $\pm$ 0.33  &  -0.42 $\pm$ 0.02  &  3.96 $\pm$ 0.03  &  5.51 $\pm$ nan  &  0.2 $\pm$ 0.27  &  -1.24 $\pm$ 0.45 \\
\hline
\hline
\label{tab:hii}
\end{longtable}
\end{center}
\end{landscape}
\twocolumn

\bsp	
\label{lastpage}
\end{document}